\documentclass[aps,showpacs,11pt]{revtex4}
\usepackage{dcolumn}
\usepackage{graphicx}
\usepackage{amsmath}
\usepackage{amsfonts}
\usepackage{amssymb}
\usepackage{psfrag}
\usepackage{wrapfig}
\usepackage{subfigure}
\usepackage{makeidx}
\usepackage{bm}
\usepackage{epsf}
\usepackage{hyperref}
\usepackage{color}
\usepackage{multirow,dcolumn}
\usepackage{graphicx}

\begin{document}

\title{Holographic Thermalization in Gauss-Bonnet Gravity with de Sitter Boundary}

\author{Shao-Jun Zhang $^{1,2}$}
\email{sjzhang84@hotmail.com}
\author{Bin Wang $^1$}
\email{wang$_$b@sjtu.edu.cn}
\author{E. Abdalla $^2$}
\email{eabdalla@usp.br}
\author{Eleftherios Papantonopoulos $^{3,4}$}
\email{lpapa@central.ntua.gr}
\affiliation{$^1$
IFSA Collaborative Innovation Center, Department of Physics and Astronomy, Shanghai Jiao Tong University, Shanghai 200240, China\\
$^2$ Instituto de
F$\acute{i}$sica, Universidade de S$\tilde{a}$o Paulo, C.P. 66318,
05315-970, S$\tilde{a}$o Paulo, SP, Brazil\\
$^3$ Department of Physics, National Technical University of Athens,
GR-15780 Athens, Greece\\
$^4$ CERN--Theory
Division, CH-1211 Geneva 23, Switzerland}

\begin{abstract}
\indent We introduce higher-derivative  Gauss-Bonnet correction
terms in the gravity sector and we relate the modified gravity
theory in the bulk to the strongly coupled quantum field theory on
a de Sitter boundary. We study the process of holographic
thermalization by examining three nonlocal observables,  the
two-point function, the Wilson loop and the holographic
entanglement entropy. We study the time evolution of these three
observables and we  find that as the strength of the Gauss-Bonnet
coupling  is increased, the saturation time of the thermalization
process to reach thermal equilibrium becomes shorter with the
dominant effect given by the holographic entanglement entropy.

\end{abstract}

\pacs{11.25.Tq, 12.38.Mh, 03.65.Ud}

\maketitle

\section{Introduction}

The AdS/CFT correspondence has been proven to be a powerful tool
in describing strongly coupled processes in quantum field theories
in regimes where  perturbation theory breaks down. This is
achieved by mapping these processes, using the holographic
principle, to processes in gravitational theories where the
coupling is  weak
\cite{Maldacena:1997re,Gubser:1998bc,Witten:1998qj}. This duality
has been applied to different areas of modern theoretical physics, including areas of
condensed matter physics like
superconductivity \cite{Horowitz:2010gk} and superfluity
\cite{Hartnoll:2009sz}.

One noticeable application of this equivalence principle is to
describe quark-gluon plasma (QGP) which is formed in heavy ion
collisions at the Relativistic Heavy Ion Collider (RHIC)
~\cite{Gelis:2011xw,Iancu:2012xa,Muller:2011ra,CasalderreySolana:2011us}.
The QGP behaves as an ideal fluid and it can be described by
finite-temperature quantum field theory \cite{QGP}. Using the
holographic principle, the hydrodynamic behavior of the thermal
field theory is identified with the hydrodynamic behavior of the
dual gravity theory. This conjecture was tested against a wide
range of thermal field theories having gravity duals
\cite{son1,son2,conductivity,fluidgravity,qnm}. Then it was found
for these field theories, the ratio of the shear viscosity to the
volume density of entropy has a universal value $ n/s = 1 / (4 \pi
)$. However, in conformal field theories dual to Einstein gravity
with curvature square corrections
\cite{Brigante:2007nu,Brigante:2008gz,Kats:2007mq} or in quantum
field theories at finite temperature having gravity duals with
hyperbolic horizons \cite{Koutsoumbas:2008wy} it was found that
the bound is violated. This deviation from the bound may indicate
that we have to go beyond the  hydrodynamic description.

In the heavy ion collisions the formation of QGP after a
characteristic time reaches a thermal equilibrium and a
hydrodynamic description can be used to understand  the
near-equilibrium physics of the QGP. However, the process to reach
thermal equilibrium, termed as the thermalization process, can not
be described by hydrodynamics. Results from RHIC show that the
time scale for equilibration of matter is considerably shorter
than expected from the use of perturbative hydrodynamic
description to thermalization \cite{Baier:2000sb,Mueller:2005un}.
This indicates that the formation of QGP is a strongly coupled
process and this motivates the use of the AdS/CFT correspondence
to study the thermalization of strongly coupled plasmas.

According to the holographic principle the gravitational dual of
the process of equilibration has to be specified. This
gravitational process will be dual to the dynamical passage of a
system from a pure state in its low-temperature phase to an
approximated thermalized state in its high temperature phase. The
proposed gravitational dual process was the black hole formation
via gravitational collapse of a scalar field in AdS space
\cite{danielsson,janik1,chesler,garfinkle,Garfinkle:2011tc}, or
 of a collapsing  thin shell of matter
described by an AdS-Vaidya metric
\cite{bhattacharya,lin,Balasubramanian:2010ce,Balasubramanian:2011ur}.

To study the detailed process of the thermalization, local
observables in quantum field theories  on the boundary, such as
the energy-momentum tensor and its derivative, are not sufficient.
We need to use some nonlocal observables to probe the
process~\cite{Balasubramanian:2010ce,Balasubramanian:2011ur}.
According to the holographic principle, in the dual gravity
description, the expectation value
  of local gauge-invariant operators is determined by the asymptotical
  behavior of the metric close to the AdS boundary. However, in a holographic thermalization process,
  the metric out of the collapsing matter is fixed during the collapse, so that the expectation value of local
  gauge-invariant operators cannot track the process. While nonlocal observables are dual to
  geometric quantities which can reach deeper into the bulk spacetime and thus provide detailed
  information about the process. Three nonlocal observables were used, the two-point
function, the Wilson loop and the entanglement entropy which can
probe different regions of the field system and reflect different
aspects of the thermalization process. These three observables can
all be evaluated by calculating some geometric quantities in the
gravity side.

An interesting extension  is to study thermalization process in
boundary quantum field theories leaving on a curved spacetime
(QFTCS). The holographic understanding of the strongly coupled
QFTCS were proposed in de Sitter (dS) spacetime
~\cite{Hirayama:2006jn,Ghoroku:2006af,Marolf:2010tg,Buchel:2013dla,Fischler:2013fba,Fischler:2014tka}.
In~\cite{Marolf:2010tg} an interesting holographic Einstein
gravity model was built to relate  the strongly coupled quantum
field theories on the dS boundary to the bulk Einstein gravity.
Employing this model, the authors of ~\cite{Fischler:2013fba}
examined the thermalization process of the dual quantum field
theories in dS background by using the holographic entanglement
entropy  as a probe. They argued that similar to flat boundary
case~\cite{Liu:2013iza,Liu:2013qca}, the whole thermalization
process can be divided into a sequence of processes:
pre-local-equilibration quadratic growth, post-local-equilibration
linear growth, memory loss and saturation. Moreover, they found
that the saturation time depends on the entanglement sphere
radius. When the radius is small, the saturation time increases
linearly as the radius increases. This is expected as the behavior
should reduce to coincide with the result of the flat boundary
case at this
time~\cite{Balasubramanian:2010ce,Balasubramanian:2011ur}.
However, when the radius is large, due to the existence of the
cosmological constant, the saturation time blows up
logarithmically as the radius approaches to the cosmological
horizon.

Further investigations of the thermalization process were
discussed. A model was proposed in \cite{galante,kundu} to include
the effect of a nonvanishing chemical potential $\mu$, which is
usually the case in real heavy ion collision processes. The effect
on thermalization  of the chemical potential in the framework of
Einstein gravity coupled to Born-Infeld nonlinear electrodynamics
was investigated in \cite{Camilo:2014npa}. Also   higher curvature
corrections \cite{chinesesGB,Li:2013cja} were introduced, angular
momentum \cite{arefeva1}, noncommutative and hyperscaling
violating geometries
\cite{chinesesNC,Alishahiha:2014cwa,lifshitz}.

In most of the studies of the thermalization process with a
collapsing thin shell of matter some universal features emerge.
First of all the thermal limit is reached after a finite time
which is a function of the geometric size of the
probe in the boundary field theory. All probes used show a slight
delay in the onset of thermalization, an apparent nonanalyticity
at the endpoint of thermalization, the transition to full thermal
equilibrium is instantaneous and these features are independent of
dimensionality. For homogeneous initial conditions the
entanglement entropy thermalizes slowest, and sets a timescale for
equilibration that saturates a causality bound
\cite{Balasubramanian:2010ce,Balasubramanian:2011ur}.

It is interesting to investigate what is the influence on the
description of the holographic thermalization process of the dual
strongly coupled of quantum field theories living on a curved
boundary, if we go beyond Einstein gravity introducing
higher-derivative terms such as the Gauss-Bonnet correction in the
gravity bulk. Also to investigate whether the universal properties
of the thermalization process of the dual field system observed in
Einstein gravity still persist in the case of the presence of the
higher-derivative terms  in the bulk. These are the main
motivations of the present work. We will use all of the three
nonlocal observables, including the two-point function, the
Wilson loop and the entanglement entropy to probe the
thermalization process and expect to disclose richer properties
which  resulted because of the presence of the higher-derivative
terms  in the bulk.

From the three observables used to probe the thermalization
process in the boundary theory the one which is giving more
information is the entanglement entropy. The reason is that
entanglement entropy is related more to the degrees of freedom of
the system so that it is more sensitive to the thermalization
process. The inclusion of a Gauss-Bonnet correction term in the
bulk will modify the formulas to compute the holographic
entanglement entropy~\cite{deBoer:2011wk,Hung:2011xb,Myers:2013lva,Dong:2013qoa,Camps:2013zua}.
In~\cite{Ryu:2006bv}, the formula for calculating the holographic
entanglement entropy in Einstein gravity was proposed and it was
stated that the entanglement entropy of the boundary quantum field
theory is equivalent to the area of a minimal surface in the bulk.
 However, when higher-derivative terms
are included the area law has to be modified including
 additional contributions coming from the intrinsic
curvature of the minimal surface. We will show that the
entanglement entropy is giving district features on the
thermalization process compared to the other two probes.

The paper is organized as follows. In Sec. II, in the Gauss-Bonnet
gravity, we will derive bulk gravity solutions with a foliation
such that the boundary metric corresponds to dS space in a given
coordinate system. We will give a vacuum AdS solution, a black
hole solution and its Vaidya-like version in the bulk. In Sec.
III, we will apply three nonlocal observables (two-point
function, Wilson loop and entanglement entropy) to study the
process of holographic thermalization in the Vaidya-like
background. The last section is devoted to a summary and
discussions.

\section{Gravity solutions with de Sitter slices}

In this section, we will discuss the bulk gravity solutions with a
Gauss-Bonnet correction term and a foliation such that the
boundary metric corresponds to a de Sitter space. We are going to
present three bulk solutions, including a vacuum AdS solution, a
static AdS black hole solution and its Vaidya-like solution.

\subsection{The action}

We consider the Gauss-Bonnet gravity theory in
$(d+1)$ dimensions ($d\geq 4$) with the action
\begin{eqnarray}\label{action}
S = \frac{1}{16\pi G_N^{(d+1)}} \int d^{d+1} x \ \sqrt{-g}
\bigg[\frac{d(d-1)}{L^2}+R + \frac{L^2 \lambda}{(d-2)(d-3)}
\left(R_{\mu\nu\rho\sigma} R^{\mu\nu\rho\sigma} - 4 R_{\mu\nu}
R^{\mu\nu} + R^2 \right)\bigg]~, \nonumber \\
\end{eqnarray}
where the first term $\frac{d(d-1)}{L^2}$
corresponds to the cosmological constant.
$\lambda$ is the Gauss-Bonnet factor, which is
usually constrained within the range~\cite{Buchel:2009tt,Camanho:2009vw,Buchel:2009sk}
\begin{eqnarray}\label{lambda-range}
-\frac{(d-2) (3d+2)}{4(d+2)^2} \leq \lambda \leq
\frac{(d-2)(d-3)(d^2-d+6)}{4(d^2-3d+6)^2}~,
\end{eqnarray}
by respecting the causality of the dual field theory on the
boundary and preserving the positivity of the energy flux in CFT
analysis. In our discussion below, we will allow $\lambda$ to go
beyond this constraint to examine how the violation of causality
can influence the thermalization and further  to study  if the
holographic thermalization process can put some constraints on
$\lambda$.

From the above action, we can derive the equations of motion
\begin{eqnarray}\label{eom}
R_{\mu\nu} -\frac{1}{2} R g_{\mu\nu} -\frac{d(d-1)}{2 L^2}
g_{\mu\nu} +\frac{L^2 \lambda}{(d-2)(d-3)} H_{\mu\nu}=0~,
\end{eqnarray}
where
\begin{eqnarray}
H_{\mu\nu}=2 \left(R_{\mu\rho\sigma\xi} R_\nu^{~\rho\sigma\xi}-2 R_{\mu\rho\nu\sigma} R^{\rho\sigma}-
2 R_{\mu\rho} R^\rho_{~\nu} +R R_{\mu\nu}\right)-\frac{1}{2} \left(R_{\alpha\beta\rho\sigma} R^{\alpha\beta\rho\sigma}
 - 4 R_{\alpha\beta} R^{\alpha\beta} + R^2\right) g_{\mu\nu}~.\nonumber \\
\end{eqnarray}

For asymptotically AdS spacetime, the metric can be written in the Fefferman-Graham form~\cite{Fefferman:1985}
\begin{eqnarray}\label{Fefferman-Graham}
ds^2 = \frac{\tilde{L}^2}{z^2} \left(g_{\mu\nu}(z,x) dx^\mu dx^\nu
+ dz^2\right)~,
\end{eqnarray}
where $\tilde{L}$ is the AdS radius which is fixed in terms of $L$
as we will see in the following. From the above form, we can read
off the metric of the dual quantum field theory, which lives at
the boundary $z=0$, as $ds^2=g_{\mu\nu}(0,x) dx^\mu dx^\nu$. In
this paper, we are interested in cases in which $g_{\mu\nu}(0,x)$
corresponds to dS space in a given system.

\subsection{An AdS vacuum solution}

The first solution we are going to discuss in this subsection is
an AdS vacuum solution. It is dual to the vacuum state of the dual
quantum field theory. For free field theory, there is a well-known
vacuum state, the Bunch-Davies or the Euclidean vacuum, which is
dS invariant and can be reduced to the standard Minkowski vacuum
in the limit $H\rightarrow 0$~\cite{Bunch:1978yq}. Here we will
consider the Bunch-Davies vacuum, which is well defined in the
entire space.

In the static patch, the dS metric is
\begin{eqnarray}\label{dS-staticpatch}
ds^2 = - (1-H^2 r^2) dt^2 +\frac{dr^2}{1-H^2 r^2} +r^2
d\Omega_{d-2}^2~,
\end{eqnarray}
which only covers one-fourth of the entire de
Sitter space. There is a cosmological horizon at
$r=1/H$ associated with a geodesic observer
sitting at $r=0$. For such an observer, the
Bunch-Davies vacuum appears to have temperature
$T_{dS}=H/2\pi$ natural for the existence of
the cosmological horizon.

To obtain a bulk solution with such boundary metric, following
\cite{Fischler:2013fba}, we write the $(d+1)$-dimensional bulk
metric in the Fefferman-Graham form,
\begin{eqnarray}\label{ads-staticpatch}
ds^2 = \frac{\tilde{L}^2}{z^2} \left(-f(r,z) dt^2 +j(r,z) dr^2
+h(r,z) d\Omega_{d-2}^2 +dz^2\right)~.
\end{eqnarray}
with
\begin{eqnarray}
f(r,0)=1-H^2 r^2,\qquad j(r,0)=\frac{1}{1-H^2 r^2}~,\qquad
h(r,0)=r^2~.
\end{eqnarray}
 The unknown functions can be determined using
perturbation methods \cite{Fischler:2013fba}, as follows
\begin{eqnarray}\label{metric_function}
f(r,z)&=&(1-H^2 r^2) \left(1-\frac{H^2 z^2}{4}\right)^2,\nonumber\\
j(r,z)&=&\frac{1}{(1-H^2 r^2)} \left(1-\frac{H^2 z^2}{4}\right)^2,\nonumber\\
h(r,z)&=&r^2 \left(1-\frac{H^2 z^2}{4}\right)^2.
\end{eqnarray}
We note that the Eqs.~(\ref{ads-staticpatch}) and
(\ref{metric_function}) are the same as the ones derived in
\cite{Fischler:2013fba} for the Einstein case except with a
modified AdS curvature scale $\tilde{L}$, which is related to the
cosmological constant by the relation $\tilde{L}^2=
\frac{1+\sqrt{1-4\lambda}}{2}L^2$. The solution has a regular
Killing horizon at $z=2/H$ with constant surface gravity and this
is in fact related to the de Sitter temperature $T_{dS} = H/2\pi$.

\subsection{A black hole solution}

Our aim is to study the holographic thermalization of the dual
field theory system on the boundary starting from a
nonequilibrium state at the beginning. The process can be
described holographically by a Vaidya-like geometry in the bulk.

The static patch of dS$_d$ can be written in the form as
\begin{eqnarray}
ds^2=(1-H^2 r^2) \left(-dt^2 + \frac{dr^2}{(1-H^2 r^2)^2}
+\frac{r^2}{1-H^2 r^2} d\Omega_{d-2}^2\right)~.
\end{eqnarray}
Defining a new coordinate $r \equiv \frac{1}{H}
\tanh \xi$, the above metric becomes
\begin{eqnarray}
ds^2 = (1-H^2 r^2) \left(-dt^2 +d \Sigma_{d-1}^2\right)~.
\end{eqnarray}
This static patch of dS$_d$ is conformally related to the
Lorentzian hyperbolic cylinder $\mathbb{R} \times
\mathbb{H}_{d-1}$, where $\mathbb{H}_{d-1}$ is the Euclidean
hyperboloid $d \Sigma_{d-1}^2 = d\xi^2 +\sinh^2 \xi
d\Omega_{d-2}^2$. A black hole solution in the bulk exist,
 with a boundary being the Lorentzian hyperbolic
cylinder $\mathbb{R} \times \mathbb{H}_{d-1}$ and it reads,
\begin{eqnarray}\label{Hbh}
ds^2 &=& -f(\rho) dt^2 +\frac{1}{f(\rho)} d\rho^2 + \rho^2 d\Sigma_{d-1}^2~,\\
f(\rho)&=&-1+\frac{\rho^2}{2 L^2 \lambda} \left(1-\sqrt{1+4
\lambda\left(\frac{m}{\rho^d}-1\right)}\right)~.\nonumber
\end{eqnarray}
This is the well-known Gauss-Bonnet black hole with
$k=-1$~\cite{Cai:2001dz} where $m$ is related to the mass of the
black hole. In the Einstein limit $\lambda \rightarrow 0$, the
above black hole solution reduces to the topological black hole
described in refs.
\cite{Emparan:1998he,Birmingham:1998nr,Emparan:1999gf}. The
horizon $\rho_+$ is given by the largest positive root of
$f(\rho)=0$. With the horizon, the mass parameter can be expressed
as
\begin{eqnarray}
m=\rho_+^{d-4} \left(\rho_+^4- L^2 \rho_+^2  +L^4 \lambda\right)~.
\end{eqnarray}
The Hawking temperature is
\begin{eqnarray}\label{temperature}
T =\frac{1}{4\pi} \frac{d}{d\rho} f(\rho)\bigg|_{\rho_+} = \frac{d
\rho_+^4 - (d-2) L^2 \rho_+^2 +(d-4) L^4 \lambda}{4\pi L^2 \rho_+
(\rho_+^2 - 2 L^2 \lambda)}~.
\end{eqnarray}
We should note that the zero temperature solution of Eq.~(\ref{Hbh}) is not the solution with $m=0$ which is isometric to
the AdS vacuum solution. Given a fixed $\lambda$, the smallest
black hole has the horizon radius
\begin{eqnarray}
\rho_{min}^2 = \frac{(d-2) L^2}{2 d} \left(1+\sqrt{1-\frac{4
d(d-4)}{(d-2)^2}\lambda}\right)~,
\end{eqnarray}
which has vanishing Hawking temperature when the
black hole mass is most "negative", that is
\begin{eqnarray}
m_{ min} = - \frac{(d-2) L^4 \rho_{min}^{d-4}}{d^2}
\left(1-\frac{4d}{d-2} \lambda + \sqrt{1-\frac{4
d(d-4)}{(d-2)^2}\lambda}\right)~.
\end{eqnarray}
This means that when the mass is negative in the
range $0>m>m_{min}$, the black hole still has
a regular horizon and reasonable thermodynamics
with $T<T_{dS}$. This is a typical
behavior of topological black holes \cite{Vanzo:1997gw}. In Fig. 1, we plot the relation
between the temperature $T$ and the mass
parameter $m$. From the figure, we can see that
the minimal mass increases as $\lambda$
increases. The three curves intersect at the
point for the AdS vacuum solution when $m=0$.

\begin{figure}[!htbp]
\centering
\includegraphics[width=0.6\textwidth]{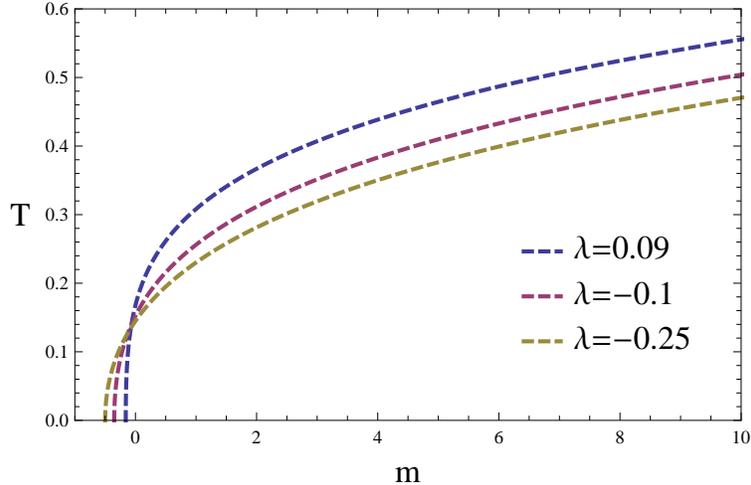}
\caption{Hawking temperature $T$ versus the mass
parameter $m$ for $d=4$. We have set $L=1$. The
three curves are for $\lambda=-0.25, -0.1, 0.09$, respectively, and they intersect at the  AdS
vacuum solution when $m=0$.}
\end{figure}

Going back to the $r$ coordinate, the metric becomes
\begin{eqnarray}\label{bulkBH}
ds^2 = \frac{\rho^2}{1-H^2 r^2} \left(-\frac{f(\rho)}{\rho^2}
(1-H^2 r^2) dt^2 +\frac{dr^2}{1-H^2 r^2}+r^2
d\Omega_{d-2}^2\right)+\frac{d\rho^2}{f(\rho)}~,
\end{eqnarray}
whose boundary metric is just (or conformal to) the dS metric in
static patch equation (\ref{dS-staticpatch}). Holographically,
this bulk geometry is dual to the thermal quantum field theorie on
the static patch of $dS_d$ at the temperature given by Eq.~
(\ref{temperature}). Note that this temperature does not have to
be the same as the dS temperature $T_{dS}$, and this will yield a
conical singularity in the manifold unless $T=T_{dS}$. We do not have to worry about this
singularity since we only have interest on the
physics outside the horizon, see more discussions
about this point in Ref.~\cite{Fischler:2013fba}.

\subsection{A Vaidya-like solution}

To study the process of the thermalization in the boundary quantum
field field theory the bulk should be  holographically modelled by
the process of a black hole formation.  Thus, we need a
Vaidya-like version of the black hole solution of Eq.~(\ref{bulkBH}). Defining an inverse radius $z=L^2/\rho$, going to
Eddington-Finkstein coordinates and introducing a time-dependent
mass parameter, we obtain the Vaidya-like bulk metric
\cite{Fischler:2013fba}
\begin{equation}\label{adsvaidya}
ds^2 = \frac{L^2}{z^2} \left(-f(z,v) dv^2-2 dv dz +\frac{H^2
L^2}{(1-H^2 r^2)^2} dr^2 +\frac{H^2 L^2}{1-H^2 r^2} r^2 d
\Omega_{d-2}^2\right)~,
\end{equation}
where the metric function in the presence of the Gauss-Bonnet term
is
\begin{equation}
 f(z,v)= \frac{z^2}{L^2}
\bigg[-1+\frac{L^2}{2 \lambda z^2} \left(1-\sqrt{1+4
\lambda\left(m(v) z^d-1\right)}\right)\bigg]~.\nonumber
\end{equation}

To obtain (\ref{adsvaidya}) the equation of motion (\ref{eom}) has
to be modified by introducing an external source,
\begin{eqnarray}
R_{\mu\nu} -\frac{1}{2} R g_{\mu\nu} -\frac{d(d-1)}{2 L^2}
g_{\mu\nu} +\frac{L^2 \lambda}{(d-2)(d-3)} H_{\mu\nu}= 8\pi G
T_{\mu\nu}^{ext}~.
\end{eqnarray}
Substituting the Vaiya-like metric (\ref{adsvaidya}) into the
above equation, one can obtain the energy-momentum tensor of the
required external source,
\begin{eqnarray}
8\pi G T_{\mu\nu}^{ext} = \frac{d-1}{2} z^{d-1} \frac{d m}{dv}
\delta_{\mu v} \delta_{\nu v}~,
\end{eqnarray}
which implies that the infalling shell is made of null dust. The
mass function $m(v)$ can take two different forms as follows
\cite{Fischler:2013fba}:
\begin{itemize}
\item $m (v) =\frac{M}{2} \left(1-\tanh \frac{v}{v_0}\right)$ with $m_{min}<M<0$

This choice is equivalent to preparing the field
system in a state with $T<T_{dS}$ and then
letting it evolve to the Bunch-Davies vacuum.

\item $m (v) =\frac{M}{2} \left(1+\tanh \frac{v}{v_0}\right)$ with $M>0$

This corresponds to a case that we start from the Bunch-Davies
vacuum and then evolve to a state with $T>T_{dS}$.
\end{itemize}

In this paper, we will focus on the second choice with the
geometric  picture that a null dust shell falls from the boundary
into the bulk to form a black hole. At the field theory side, it
corresponds to a sudden injection of energy into the system and
then let it evolves to thermal equilibrium.

\section{Nonlocal observables}

In this section, we use three nonlocal observables--equal-time
two-point function, Wilson loop and entanglement entropy--to probe
the thermalization process. According to the holographic
principle, these three probes in the saddle approximation
correspond to some geometric quantities in the dual bulk geometry
with a metric given by (\ref{adsvaidya}). It is expected that
these observables provide insights into the thermalization process
in the strongly coupled quantum field theories  the boundary. We
will study the time evolution behavior of these three observables
after a quantum quench, and examine the influence of the
Gauss-Bonnet factor and the spacetime dimensions in the
thermalization process. Relying on numerical calculations and
without loss of generality following
\cite{Balasubramanian:2010ce,Balasubramanian:2011ur}, we will
consider a thin shell with a small $v_0$, i.e., $v_0=0.01$, and we
will set $L=1$ and fix the mass parameter to $M=1$.

\subsection{Two-point function}

Defining $\tilde{r} \equiv Hr$ and choosing two antipode points on
the sphere $\tilde{r}=\tilde{R}$ on the boundary at the boundary
time $\tilde{t}$,  we can calculate the equal-time correlation
function of some operator with large conformal dimension, which
depends on $\tilde{t}$. Holographically, in the saddle
approximation this two-point function corresponds to the length of
the geodesic in the bulk which is anchored at the two points on
the boundary. The geodesic $\gamma$ can be parametrized by two
functions, $z(\tilde{r})$ and $v(\tilde{r})$, with other
coordinates fixed respecting the spherical symmetry. With the
Vaidya-like metric (\ref{adsvaidya}), we can obtain the induced
metric on the geodesic, which is
\begin{eqnarray}
ds_{\gamma}^2 = \frac{L^2}{z^2}
\left(\frac{L^2}{(1-\tilde{r}^2)^2}-f(z,v) v'^2 - 2 v' z'\right)
d\tilde{r}^2~.
\end{eqnarray}
Then, the length functional of the geodesic is
\begin{eqnarray}\label{lengthfunctional}
{\cal L} &=& 2 L \int_0^{\tilde{R}} \frac{d\tilde{r}}{z} Q ~,\\
Q &\equiv&  \sqrt{\frac{L^2}{(1-\tilde{r}^2)^2}-f(z,v) v'^2 - 2 v'
z'}~.\nonumber
\end{eqnarray}
To minimize the length of the geodesic, we need to solve the two
equations of motion for $z(\tilde{r})$ and $v(\tilde{r})$, respectively, which are derived varying the length functional
(\ref{lengthfunctional}),
\begin{eqnarray}
&&2 (\tilde{r}^2-1)^4 z v'^2 z'' +\Bigg\{\left[6 L^2 (\tilde{r}^2-1)^2 v' - 6(\tilde{r}^2-1)^4 f v'^3 +(\tilde{r}^2-1)^4 z v' \left(3 v'^2 \frac{\partial f}{\partial z}-2 v''\right)\right] z'\nonumber\\
&&+ (\tilde{r}^2-1) \left[4L^2 \tilde{r} v' + 2L^2 (\tilde{r}^2-1) v''-L^2 (\tilde{r}^2-1) v'^2 \frac{\partial f}{\partial z} + (\tilde{r}^2 -1)^3 v'^4 \left(\frac{\partial f}{\partial v}+f \frac{\partial f}{\partial z}\right)\right] z\nonumber\\
&&- 2 (\tilde{r}^2-1)^4 f^2 v'^4 -4(\tilde{r}^2-1)^4 v'^2 z'^2 +4 L^2 (\tilde{r}^2-1)^2 f v'^2 -2L^4\Bigg\}=0~,\label{lengthEoM1}\\
&&2 (\tilde{r}^2 -1) (L^2 f +(\tilde{r}^2-1)^2 z'^2) z v'' +\Bigg\{(\tilde{r}^2-1)^3 z' \left[2 f^2 -z \frac{\partial f}{\partial v}- z f \frac{\partial f}{\partial z}\right] v'^3\nonumber\\
&&+\left[6 (\tilde{r}^2-1)^3 f z'^2 +L^2 (\tilde{r}^2-1)z\frac{\partial f}{\partial v} -3(\tilde{r}^2-1)^3 z z'^2 \frac{\partial f}{\partial z}\right] v'^2\nonumber\\
&&+\left[4L^2 \tilde{r} f z -2L^2 (\tilde{r}^2-1) f z' +4(\tilde{r}^2-1)^3 z'^3 -2(\tilde{r}^2-1) z z' \left((\tilde{r}^2-1)^2 z''-L^2 \frac{\partial f}{\partial z}\right)\right] v'\nonumber\\
&&+ 2L^2 \left[z (2 \tilde{r}
z'+(r^2-1)z'')-(\tilde{r}^2-1)z'^2\right]\Bigg\}=0~.\label{lengthEoM2}
\end{eqnarray}
Here prime denotes the derivative with respect to
$\tilde{r}$, i.e., $'\equiv
\frac{d}{d\tilde{r}}$. To solve these equations,
we need to  consider the symmetry of the geodesic
and impose the following boundary conditions
\begin{eqnarray}\label{boundarycondition}
&& z(\epsilon)=z_\ast+{\cal O}(\epsilon^2)~, \quad z'(\epsilon)=0+{\cal O}(\epsilon^2)~,\nonumber\\
&& v(\epsilon)=v_\ast+{\cal O}(\epsilon^2)~, \quad
v'(\epsilon)=0+{\cal O}(\epsilon^2)~,
\end{eqnarray}
where $\epsilon$ is a small quantity, with order of $10^{-3}$
typically. To avoid the numerical problem at $\tilde{r}=0$ (note
that $\tilde{r}=0$ is a singular point of Eq.~(\ref{lengthEoM1}) as $v'(0)=0$ respecting the symmetry), we
impose the boundary conditions at the neighborhood of the midpoint
$\tilde{r}=0$, rather than exactly at the midpoint. The ${\cal O}
(\epsilon^2)$ terms above are the correction terms, and can be
obtained by solving the two equations of motion around
$\tilde{r}=0$ order by order. Here, we only calculate the
equations to the order $\epsilon^2$. The two free parameters
$z_\ast$ and $v_\ast$ are determined by the constraint equations
\begin{eqnarray}
z(\tilde{R})=z_0~,\quad v(\tilde{R})=\tilde{t}~,
\end{eqnarray}
where $z_0$ is a UV cutoff and $\tilde{t}$ is the boundary time.

Using numerical methods, we can calculate the
geodesic length ${\cal L}$ at any given time,
which is cutoff dependent and divergent as $z_0
\rightarrow 0$. To remove its dependence on the
cutoff, one can define a relative geodesics
length $\delta \bar{\cal L} \equiv \frac{{\cal
L}-{\cal L}_{thermal}}{R_{rs}}$, with ${\cal
L}_{thermal}$ being the length of the late time
and $R_{rs}=L \ln
\left(\frac{1+\tilde{R}}{1-\tilde{R}}\right)$ is
the proper radial separation between the two
points on the boundary. $\delta \bar{\cal L}$ is
a function of the boundary time $\tilde{t}$ and
$\tilde{R}$ is the boundary separation.

In Fig.~2, we plot the time evolution of $\delta \bar{\cal L}$ in
five dimensions ($d=4$) for two boundary scales $\tilde{R}=0.3$
and $\tilde{R}=0.9$. We take the Gauss-Bonnet factor in a range
bigger than the constraint (2) by choosing $\lambda = -0.25,
-\frac{7}{36}(\text{lower bound}), -0.1, 0, 0.03,
0.09(\text{upper bound}), 0.2$, respectively. In Fig. 2 we can see
the whole thermalization process. At very early time the evolution
encounters a delay, especially for large $\tilde{R}$, then it
follows a pre-local-equilibration stage during which the growth is
quadratic in time; later, it appears in a post-local-equilibration
stage of linear growth, and finally there emerges a period of memory
loss prior to equilibration, after which the curves flatten out
and the two-point functions reach their thermal equilibrium
values.

For large $\tilde{R}$, it takes more time for the two-point
function to reach thermal equilibrium value. This can be
understood by the fact that the collapsing shell from the boundary
passes the geodesic with larger length slower than the one with
smaller length. These phenomena, disclosed in Fig. 2, are
consistent with that observed in the flat boundary
case~\cite{Liu:2013iza,Liu:2013qca,Balasubramanian:2010ce,Balasubramanian:2011ur}
and also in the Einstein dS boundary case~\cite{Fischler:2013fba}.
They are universal and do not change in case that the high
curvature corrections are considered in the gravity, even when the
Gauss-Bonnet coupling constant $\lambda$ takes values beyond the
constraint appearing in Eq.~ (\ref{lambda-range}).

\begin{figure}[!htbp]
\centering
\includegraphics[width=0.3\textwidth]{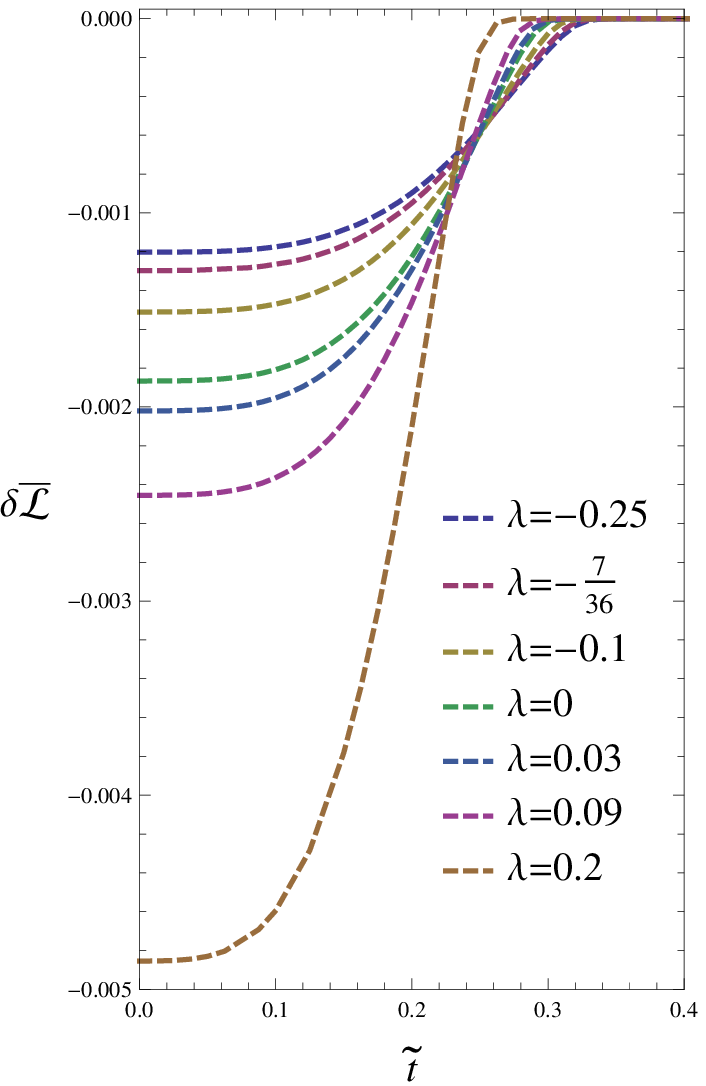}\qquad\qquad
\includegraphics[width=0.29\textwidth]{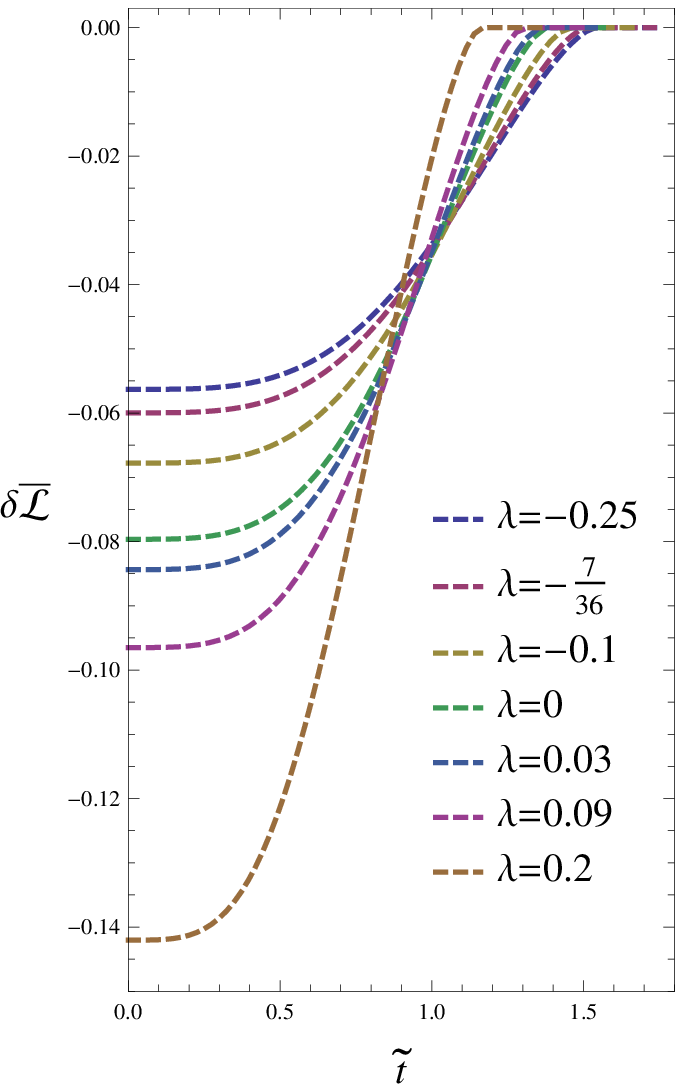}
\caption{Time evolution of $\delta \bar{\cal L}$ in five
dimensions ($d=4$) for the two boundary scale $\tilde{R}=0.3,  0.9$,
respectively. In each graph, on the left, from top to bottom,
seven curves take Gauss-Bonnet coupling constant $\lambda=-0.25,
-\frac{7}{36} (\text{lower bound}), -0.1, 0, 0.03, 0.09
(\text{upper bound}), 0.2$, respectively.}
\end{figure}

\begin{figure}[!htbp]
\centering
\includegraphics[width=0.3\textwidth]{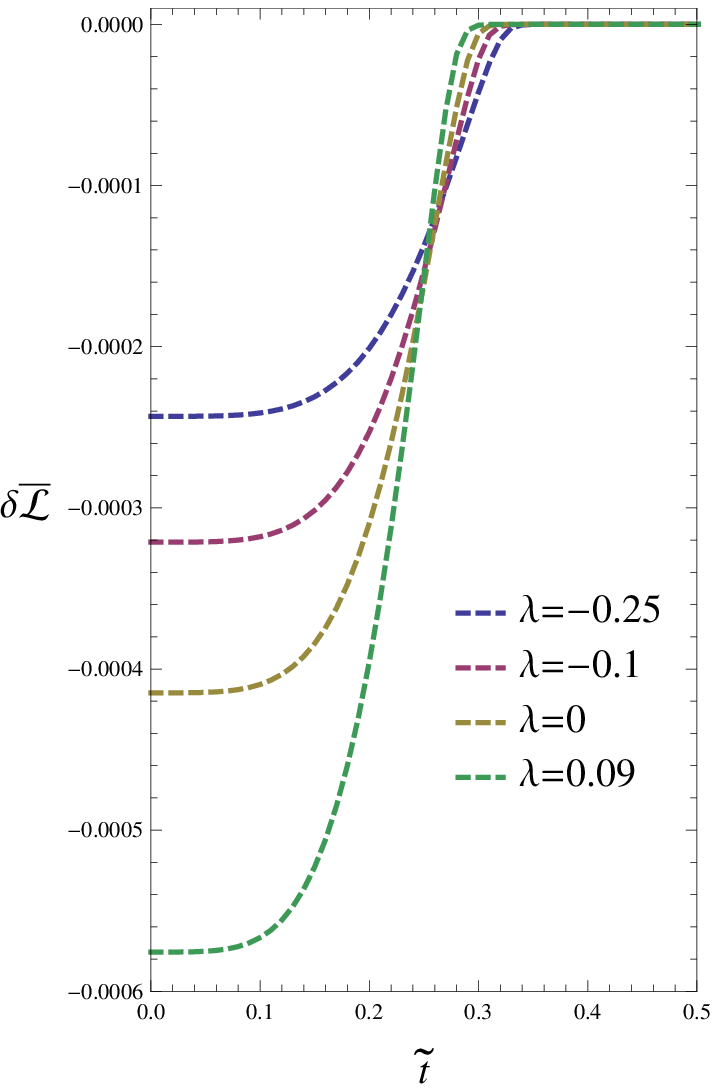}\qquad\qquad
\includegraphics[width=0.285\textwidth]{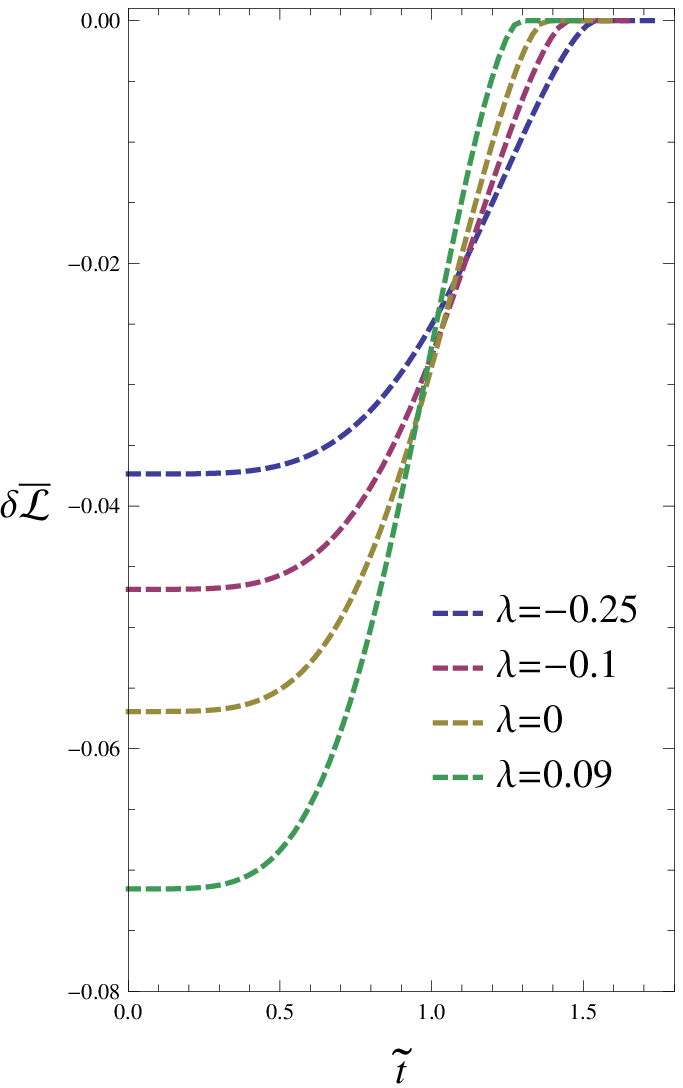}
\caption{Time evolution of $\delta \bar{\cal L}$ in six dimensions
($d=5$) with the boundary scale $\tilde{R}=0.3,  0.9$,
respectively. In each graph, on the left, from top to bottom, four
curves take Gauss-Bonnet coupling constant $\lambda=-0.25, -0.1,
0, 0.09$, respectively.}
\end{figure}

Although the Gauss-Bonnet factor does not alter qualitatively in
the five successive processes of the thermalization, in Fig.~2 we
do clearly see that it leaves imprints. The increase of the
Gauss-Bonnet factor makes the initial absolute value of $\delta
\bar{\cal L}$ to increase, which means that the dual field system
is initially further away from the thermal equilibrium. However,
the larger $\lambda$  makes the delay time shorter and the growth
of $\delta \bar{\cal L}$ faster, so that the saturation time to
reach thermal equilibrium decreases. The same phenomenon is also
observed in higher dimensions, for example as shown in Fig. 3 with
$d=5$. Comparing the results of the $d=4$ and $d=5$ cases, one can see
that, for the fixed $\lambda$, the absolute initial value of
$\delta \bar{\cal L}$ decreases as the spacetime dimension
increases, while the delay time increases, which leaves the
saturation time nearly unchanged. The dimensional influence can be
seen more clearly in Fig. 4, where we exhibit the results of two
fixed Gauss-Bonnet factors for different spacetime dimensions
($d=4$ and $d=5$).

\begin{figure}[!htbp]
\centering
\includegraphics[width=0.3\textwidth]{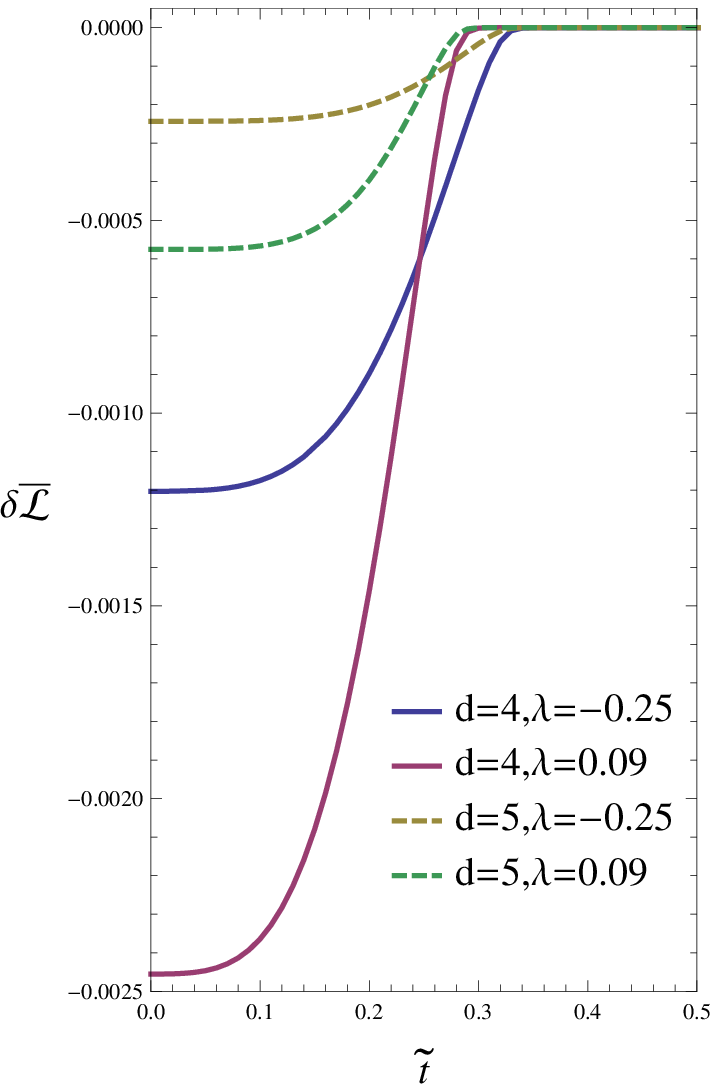}\qquad\qquad
\includegraphics[width=0.285\textwidth]{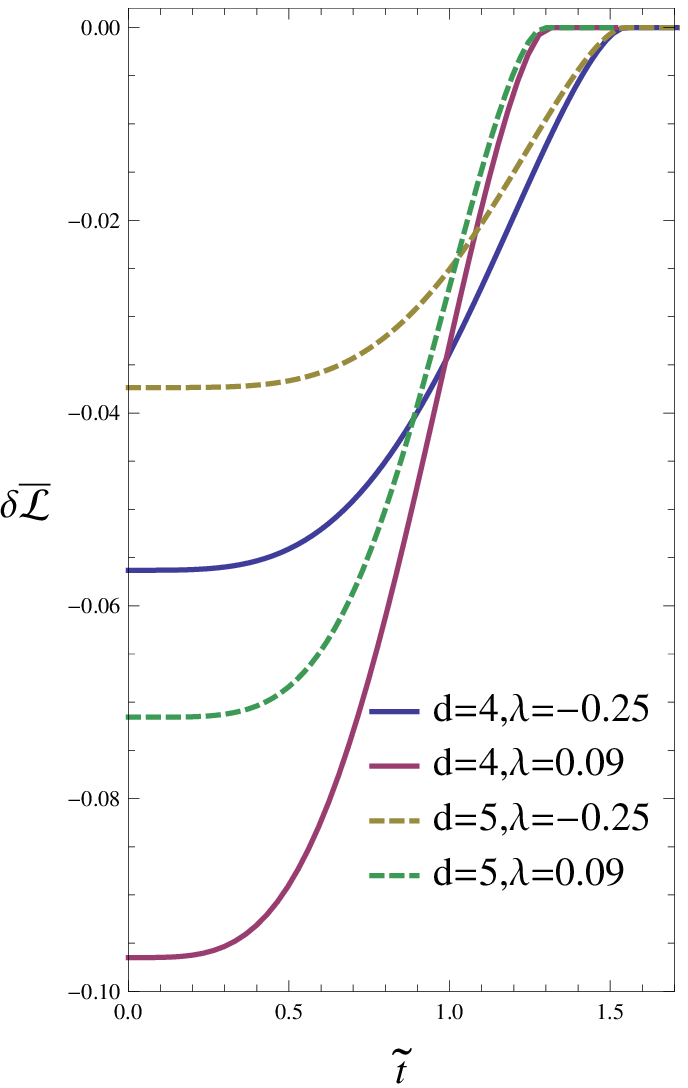}
\caption{Time evolution of $\delta \bar{\cal L}$ in different
dimensions ($d=4$ and $d=5$) with  the boundary scale
$\tilde{R}=0.3, 0.9$, respectively.}
\end{figure}

\begin{figure}[!htbp]
\centering
\includegraphics[width=0.49\textwidth]{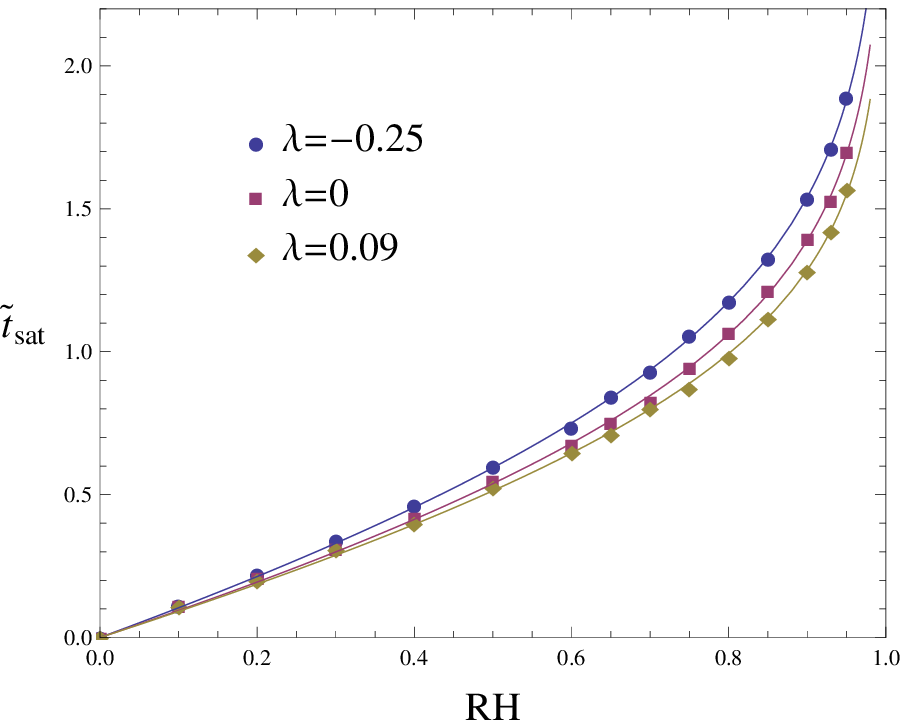}
\includegraphics[width=0.49\textwidth]{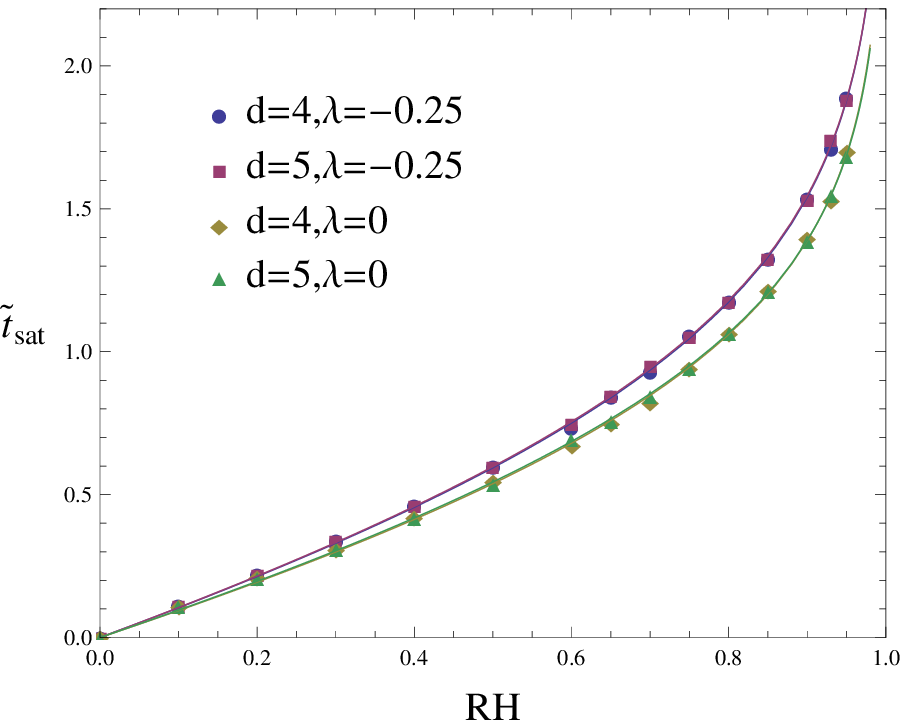}
\caption{Saturation time $\tilde{t}_{sat}$ as a
function of the boundary scale $\tilde{R}$. In
the left plot we fix the spacetime dimension
$d=4$. In the right panel, we compare the
dimensional influence by choosing two values of
the Gauss-Bonnet factor, $\lambda=-0.25$ and $\lambda=0$. Solid curves in both panels are produced by fitting the numerical points with function $a \tilde{R} + b \ln (1-\tilde{R})$.}
\end{figure}

We find that the saturation time $\tilde{t}$ is a key quantity to
characterize the thermalization process: it decreases when the
Gauss-Bonnet coupling constant increases, while it weakly depends
on the spacetime dimensions.

From Figs.~2--4 we can further learn that the saturation time
strongly depends on the boundary scale $\tilde{R}\equiv RH$. In
Fig.~5, we plot the saturation time versus the boundary scale
$\tilde{R}$. For small $\tilde{R}$, $\tilde{t}_{sat}$ increases
linearly with $\tilde{R}$ and weakly depends on the Gauss-Bonnet
constant, since the behavior should reduce to that of the flat
boundary case
there~\cite{Balasubramanian:2010ce,Balasubramanian:2011ur,Zeng:2013mca,Li:2013cja}.
However, as $\tilde{R}$ approaches the cosmological horizon,
$\tilde{t}_{sat}$ blows up logarithmically and shows strong
dependence on the Gauss-Bonnet factor. Moreover, from the right
panel of Fig. 5 one can observe that there is not a clear
influence on the saturation time measured by the two-point
function which is due to the dimensionality of spacetime.  As we
will discuss in the next subsection the same behavior is observed
also in the Wilson loop. To see an effect we have to study the
entanglement entropy as the probe.

\subsection{Wilson loop}

In this subsection, we study the time
evolution of the observable of the Wilson loop. On
the boundary at time $\tilde{t}$, we choose an equator of a sphere
with radius $\tilde{r} = \tilde{R}$, parametrized
by the angular coordinate $\theta$, and calculate
the expectation value of this Wilson loop, which depends on $\tilde{t}$. In the
bulk, it corresponds to the area of a
two-dimensional extremal surface $\Sigma$
anchored at the circle on the boundary.
Considering the symmetry, the extremal surface
$\Sigma$ can be parametrized by two functions,
$z(\tilde{r})$ and $v(\tilde{r})$, and the
induced metric on $\Sigma$ is
\begin{eqnarray}
ds_{\Sigma}^2 = \frac{L^2}{z^2}
\left(\frac{L^2}{(1-\tilde{r}^2)^2}-f(z,v) v'^2 - 2 v' z'\right)
d\tilde{r}^2 + \frac{L^4}{z^2} \frac{\tilde{r}^2}{1-\tilde{r}^2}
d\theta^2~.
\end{eqnarray}
The area functional is given by
\begin{eqnarray}\label{areafunctional}
{\cal A} &=& 2 \pi L^3 \int_0^{\tilde{R}} \frac{d\tilde{r}}{z^2} Q P~.\\
Q &\equiv& \sqrt{\frac{L^2}{(1-\tilde{r}^2)^2}-f(z,v) v'^2 - 2 v'
z'}~, \qquad P \equiv
\frac{\tilde{r}}{\sqrt{1-\tilde{r}^2}}~.\nonumber
\end{eqnarray}
To deduce the extreme value of this area
functional, we need to solve the two equations of
motion, which can be derived by varying the area
functional
\begin{eqnarray}\label{areaEoM}
&&2 \tilde{r} (\tilde{r}^2-1)^4 z v'^2 z''+ \Bigg\{-4\tilde{r} \left(L^2-(\tilde{r}^2-1)^2 f v'^2\right)^2-8\tilde{r} (\tilde{r}^2-1)^4 v'^2 z'^2\nonumber\\
&&+\left[12 \tilde{r}(\tilde{r}^2-1)^2 v' (L^2-(\tilde{r}^2-1)^2 f v'^2)+(\tilde{r}^2-1)^3 z v' \left(4v'-2\tilde{r} (\tilde{r}^2-1)v''+3\tilde{r} (\tilde{r}^2-1)v'^2 \frac{\partial f}{\partial z}\right)\right] z'\nonumber\\
&&+ (\tilde{r}^2-1) \bigg[2L^2 (2\tilde{r}^2-1)v' + 2(\tilde{r}^2-1)^2 f v'^3 +2L^2 \tilde{r}(\tilde{r}^2-1)v''-L^2 \tilde{r} (\tilde{r}^2-1)v'^2 \frac{\partial f}{\partial z} \nonumber\\
&&+\tilde{r}(\tilde{r}^2-1)^3 v'^4 \left(\frac{\partial f}{\partial v}+f \frac{\partial f}{\partial z}\right)\bigg] z\Bigg\}=0~,\\
&&2 \tilde{r} (\tilde{r}^2-1) z\left(L^2 f +(\tilde{r}^2-1)^2 z'^2\right)v''+ \Bigg\{2L^2\left[z\left((2\tilde{r}^2-1)z'+\tilde{r}(\tilde{r}^2-1)z''\right)-2\tilde{r}(\tilde{r}^2-1)z'^2\right]\nonumber\\
&&+(\tilde{r}^2-1)^2 \left[2f^2 \left(z+2\tilde{r}(\tilde{r}^2-1)z'\right)-\tilde{r}(\tilde{r}^2-1)z z' \left(\frac{\partial f}{\partial v}+ f \frac{\partial f}{\partial z}\right)\right] v'^3\nonumber\\
&&+ (\tilde{r}^2-1) \left[6(\tilde{r}^2-1)f z'\left(z+2\tilde{r}(\tilde{r}^2-1)z'\right)+\tilde{r} z \left(L^2\frac{\partial f}{\partial v}-3(\tilde{r}^2-1)^2 z'^2 \frac{\partial f}{\partial z}\right)\right] v'^2\nonumber\\
&&+\bigg[2L^2(2r^2-1)f z-4L^2 r(r^2-1)f z'+4(r^2-1)^2 z z'^2 +8r (r^2-1)^3 z'^3\nonumber\\
&&- 2\tilde{r} (\tilde{r}^2-1) z z' \left((\tilde{r}^2-1)^2
z''-L^2 \frac{\partial f}{\partial z}\right)\bigg] v' \Bigg\}=0~.
\end{eqnarray}
Again, to avoid the numerical problem at the midpoint
$\tilde{r}=0$, we impose the boundary condition
(\ref{boundarycondition}) at the neighborhood of the midpoint.

\begin{figure}[!htbp]
\centering
\includegraphics[width=0.3\textwidth]{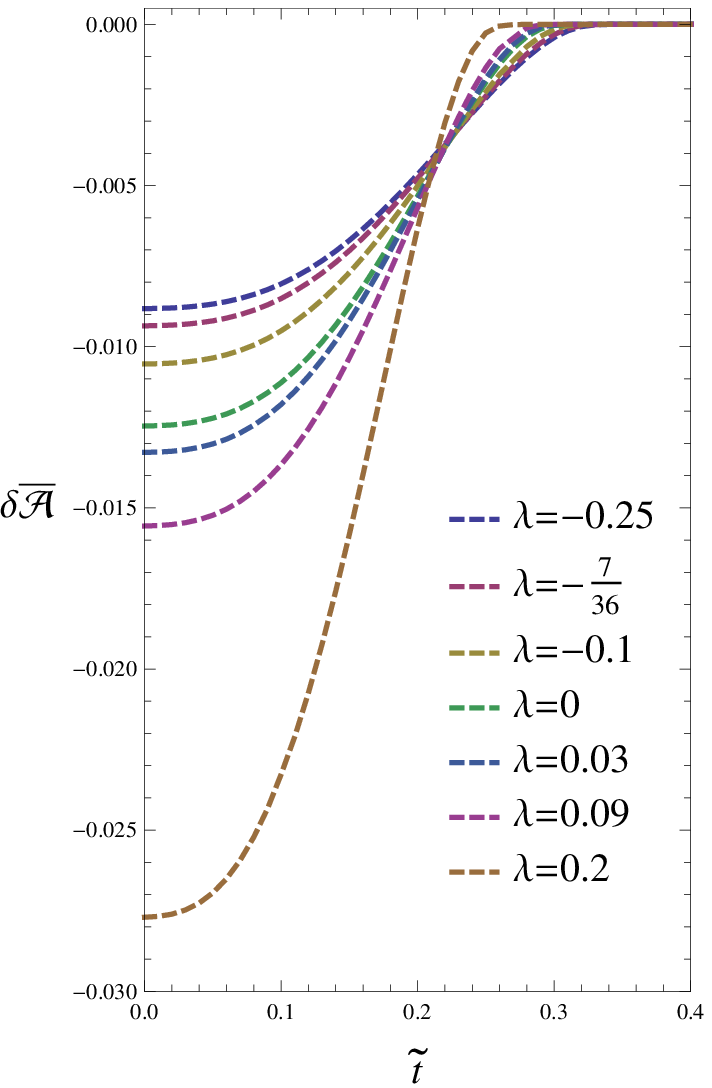}\qquad\qquad
\includegraphics[width=0.29\textwidth]{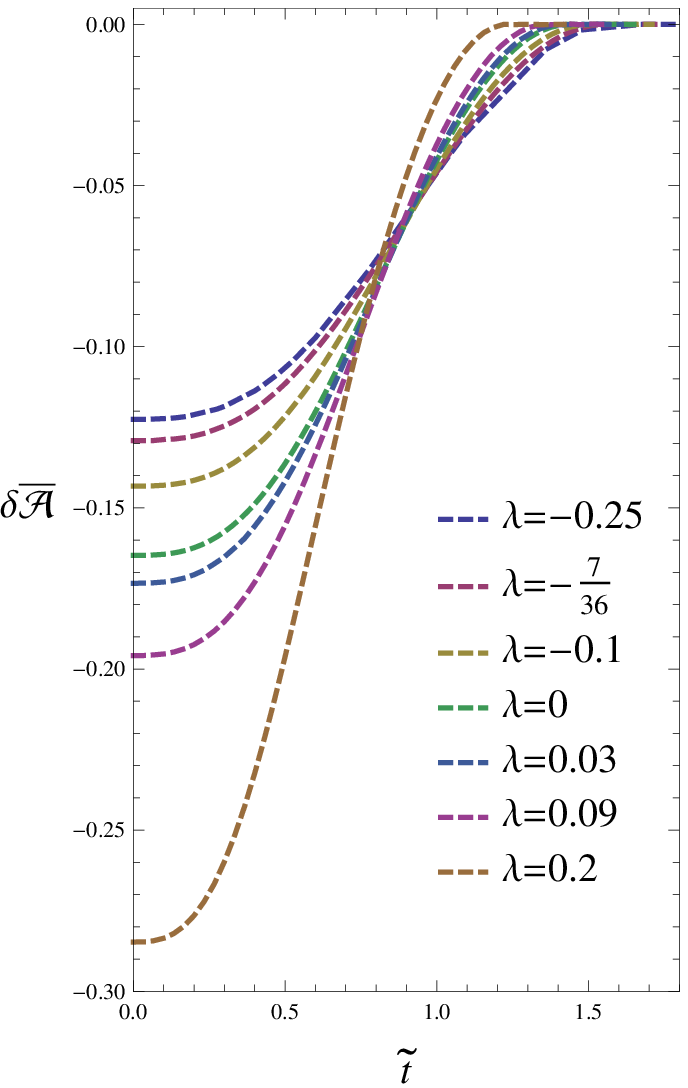}
\caption{Time evolution of $\delta \bar{\cal A}$ in five
dimensions ($d=4$) with the boundary scale $\tilde{R}=0.3,  0.9$,
respectively. In each graph, on the left, from top to bottom,
seven curves take Gauss-Bonnet coupling constant $\lambda=-0.25,
-\frac{7}{36} (\text{lower bound}), -0.1, 0, 0.03, 0.09
(\text{upper bound}), 0.2$, respectively.}
\end{figure}

\begin{figure}[!htbp]
\centering
\includegraphics[width=0.3\textwidth]{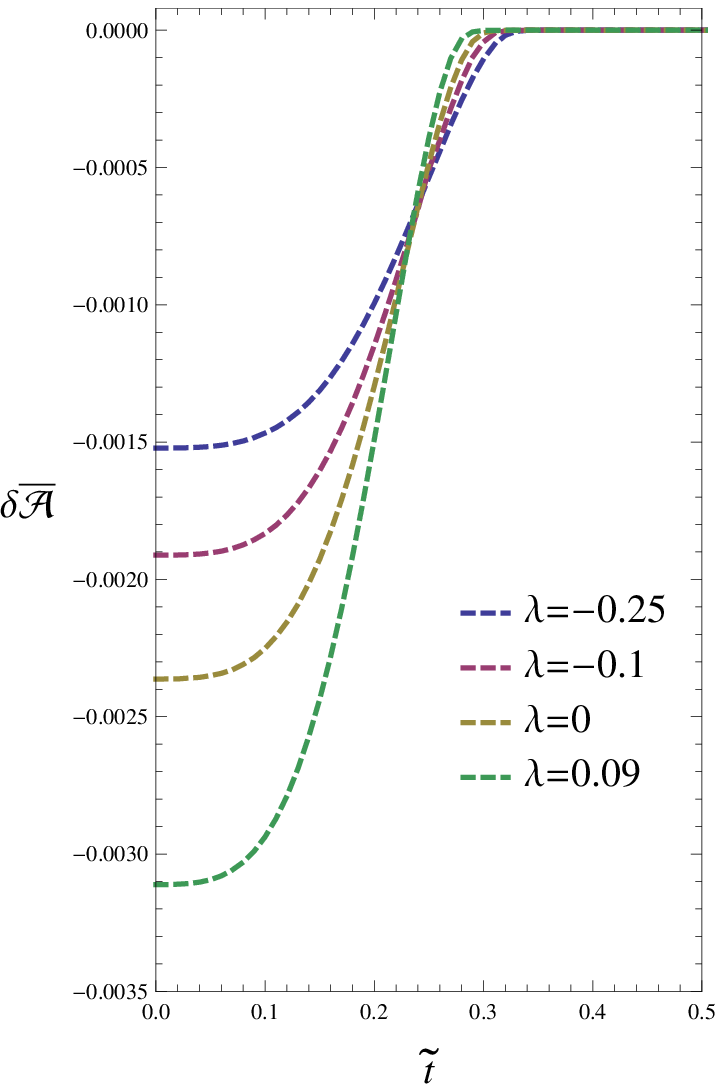}\qquad\qquad
\includegraphics[width=0.29\textwidth]{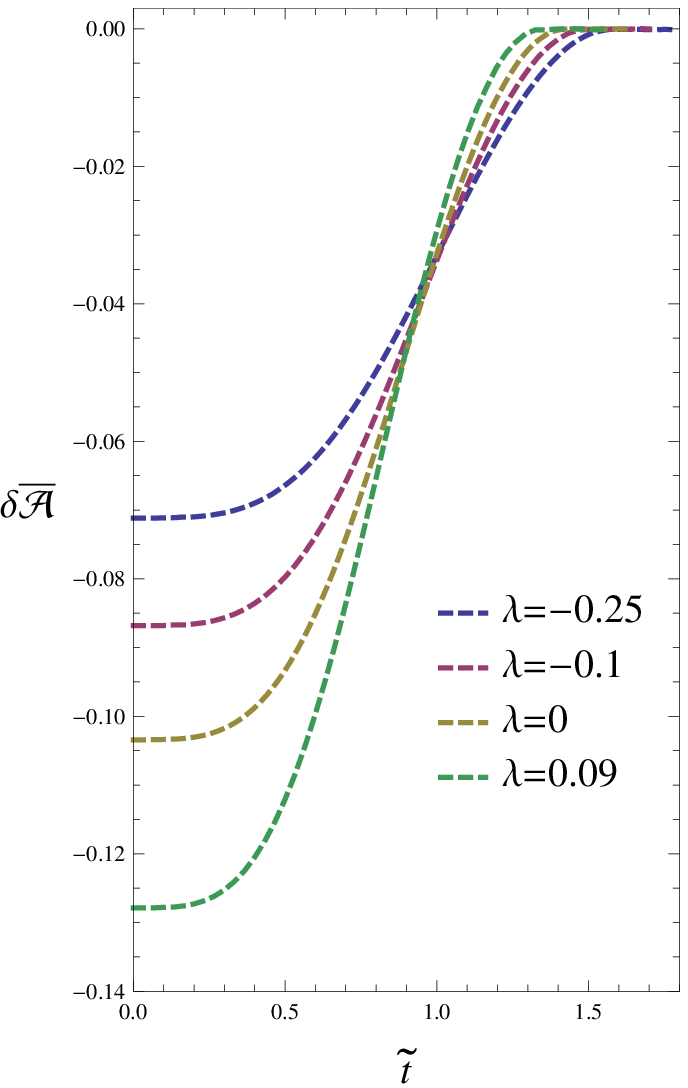}
\caption{Time evolution of $\delta \bar{\cal A}$ in six dimensions
($d=5$) with the boundary scale $\tilde{R}=0.3,  0.9$,
respectively. In each graph, on the left, from top to bottom, four
curves take Gauss-Bonnet coupling constant $\lambda=-0.25, -0.1,
0, 0.09$, respectively.}
\end{figure}

Now we study the time evolution of the related area $\delta
\bar{\cal A} \equiv \frac{{\cal A}-{\cal A}_{thermal}}{A_{bny}}$,
where $A_{bny} = 2\pi L^2
\left(\frac{1}{\sqrt{1-\tilde{R}^2}}-1\right)$ is the area of the
disk bounded by the loop on the boundary. We list the results in
Figs.~6-9, which present  similar influences of the Gauss-Bonnet
coupling and the spacetime dimensions  to those observed in the
two-point function. Fig.~9 shows the saturation time
$\tilde{t}_{sat}$, which shows the same behavior as in  Fig.~5 of
the two point function. For big boundary scale, the saturation
time deduced from the Wilson loop exhibits the consistent
behaviors with those from the two-point function, it decreases as
the Gauss-Bonnet factor increases, but it depends very mildly on
the spacetime dimensions.

\begin{figure}[!htbp]
\centering
\includegraphics[width=0.3\textwidth]{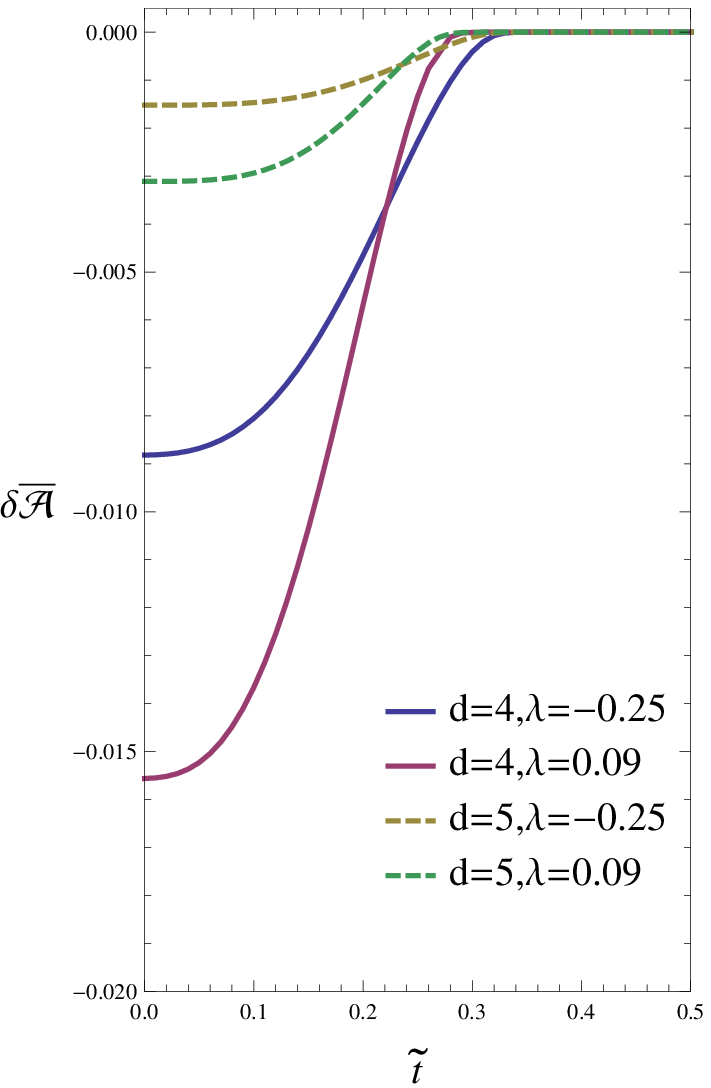}\qquad\qquad
\includegraphics[width=0.288\textwidth]{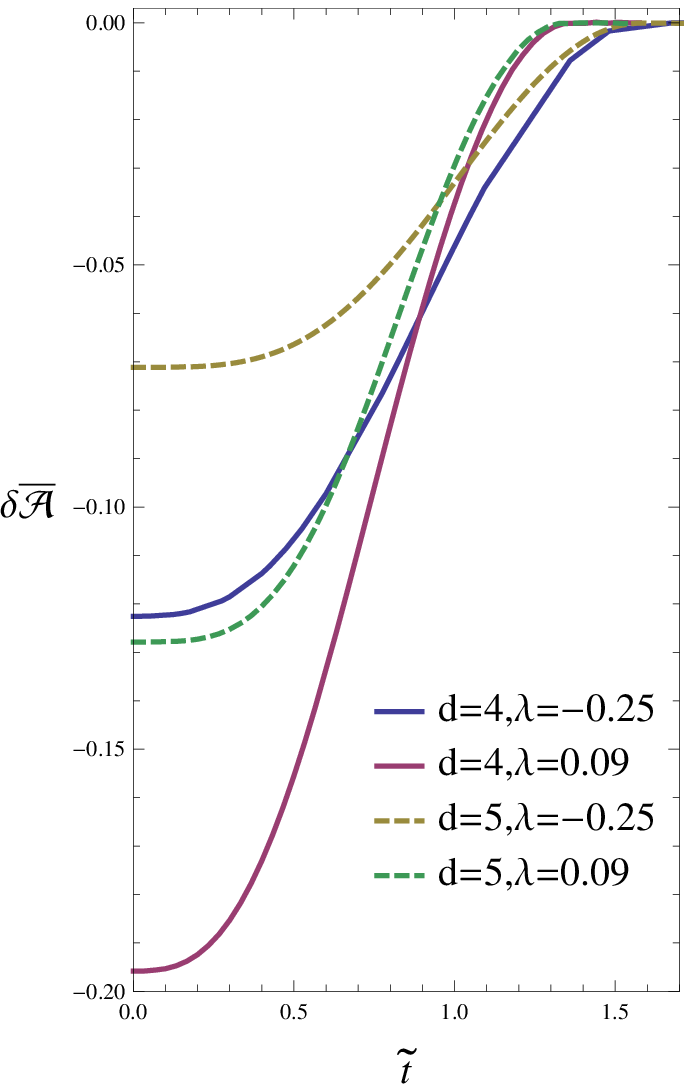}
\caption{Time evolution of $\delta \bar{\cal A}$ in different
dimensions ($d=4$ and $d=5$) and with  the boundary scale
$\tilde{R}=0.3, 0.9$, respectively.}
\end{figure}

\begin{figure}[!htbp]
\centering
\includegraphics[width=0.49\textwidth]{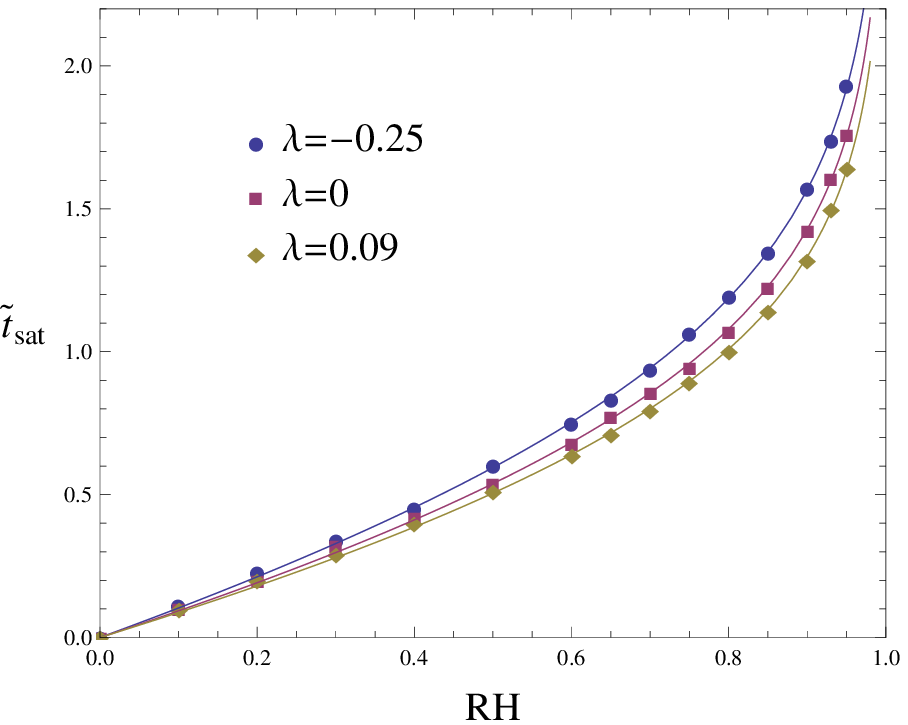}
\includegraphics[width=0.49\textwidth]{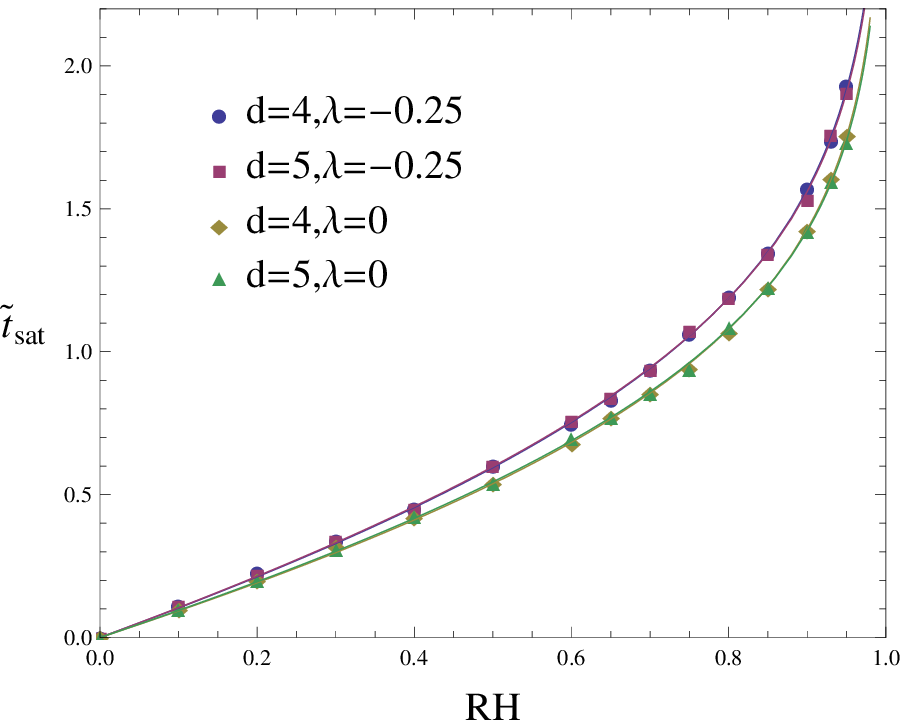}
\caption{Saturation time $\tilde{t}_{sat}$ as a
function of the boundary scale $\tilde{R}$. The
left plot has fixed spacetime dimension $d=4$. In
the right panel the comparison of the spacetime
dimensional influence for chosen Gauss-Bonnet
factors has been given. Solid curves in both panels are produced by fitting the numerical points with function $a \tilde{R} + b \ln (1-\tilde{R})$.}
\end{figure}

\subsection{Entanglement Entropy}

\subsubsection{Holographic Entanglement Entropy Formula with Gauss-Bonnet Correction}

Now we move on to study the observable of the entanglement entropy
to see how it can be used to measure the thermalization process.
Suppose the system of the quantum field theory in the boundary dS
space is divided into two parts $A$ and $A^c$, where $A^c$ is the
complement of $A$. Then we can define entanglement entropy of the
region $A$ as the Von Neuman entropy,
\begin{eqnarray}
S_A = -{\mathrm tr_A} \rho_A \log \rho_A~,
\end{eqnarray}
where $\rho_A$ is the reduced density matrix of $A$.

Generally, it is hard to calculate the entanglement entropy from
quantum field theory directly. However, AdS/CFT provides a
powerful tool to deal with this problem. A holographic
entanglement entropy has been proposed to relate this entropy to
some geometric quantity of the dual bulk geometry. In Einstein
gravity, the holographic entanglement entropy formula
is~\cite{Ryu:2006bv,Hubeny:2007xt}
\begin{eqnarray}
S_A = \frac{1}{4 G_N^{(d+1)}} {\rm ext} [{\rm Area} (\gamma_A)]~,
\end{eqnarray}
where $\gamma_A$ is the extremal codimensional
two surface in the bulk with the boundary
$\partial \gamma_A=\partial A$.

For higher derivative gravity theory, such as the
Gauss-Bonnet gravity we are considering, the
holographic entanglement formula is modified to
be~\cite{deBoer:2011wk,Hung:2011xb,Myers:2013lva,Dong:2013qoa,Camps:2013zua}
\begin{eqnarray}
S_{\rm HEE} = \frac{1}{4 G_N^{(d+1)}} \int_{\gamma_A}
\sqrt{\gamma} \left[1+ \frac{2 L^2 \lambda}{(d-2)(d-3)} {\cal
R}_{\gamma_A}\right] + \frac{1}{G_N^{(d+1)}} \frac{L^2
\lambda}{(d-2) (d-3)} \int_{\partial \gamma_A} \sqrt{h} \ {\cal
K}~.
\end{eqnarray}
The integration is done on the extremal surface
${\gamma_A}$. ${\cal R}_{\gamma_A}$ is the
intrinsic curvature scalar of $\gamma_A$, which
is the contribution due to the Gauss-Bonnet term.
The last term is added to make the variation
problem well defined. $h$ is the determinant of
the induced metric on $\partial \gamma_A$, and
${\cal K}$ is the trace of the extrinsic
curvature of $\partial \gamma_A$.

For a warped geometry with the following metric
\begin{eqnarray}\label{warpedform}
ds^2 = ds_X^2 + \sum_i e^{2 F_i(X)} ds_{Y_i}^2~,
\end{eqnarray}
we have the formulae \cite{Hung:2011xb}
\begin{eqnarray}\label{Rformulae}
R=R^X + \sum_i \left[e^{-2 F_i} R^{Y_i}-2d_i (\nabla^2 F_i) -d_i
(\partial F_i)^2\right]-\sum_{i,j} d_i d_j (\partial F_i \cdot
\partial F_j)~,
\end{eqnarray}
where $X$ and $Y_i$ are subspaces of the manifold and $d_i$ is the
dimension of $Y_i$. $F_i (X)$ is the warped factor which depends
only on the coordinates of $X$. All derivatives are evaluated in
the $X$ space and $\nabla$ denotes the covariant derivative
compatible with metric on $X$. In this subsection, the number of
dimension of $X$ space is only $1$ and then $R^X=0$, and the space
$Y$ is a $(d-2)$-dimensional sphere. Calculating the holographic
entanglement entropy by using the formula (\ref{Rformulae}), we
have
\begin{eqnarray}\label{HEE}
S_{\rm HEE}=\frac{1}{4 G_N^{(d+1)}} \Omega_{d-2} \int dX \
\sqrt{h_{XX}} e^{(d-2) F} \left[1+2 L^2 \lambda \left(e^{-2F}
+h^{XX} F'^2\right)\right]~.
\end{eqnarray}

\subsubsection{Time Evolution of Holographic Entanglement Entropy}

On the boundary at time $\tilde{t}$, we choose the entangled region
$A$ to be a $(d-1)$-dimensional sphere with
$\tilde{r}=0$ as the original point. Considering
the symmetry, the extremal codimensional two
surface $\gamma_A$ in the bulk can be
parametrized by functions $z(\tilde{r})$ and
$v(\tilde{r})$. The induced metric on $\gamma_A$
is
\begin{eqnarray}
ds_{\Sigma}^2 = \frac{L^2}{z^2}
\left(\frac{L^2}{(1-\tilde{r}^2)^2}-f(z,v) v'^2 - 2 v' z'\right)
d\tilde{r}^2 + \frac{L^4}{z^2} \frac{\tilde{r}^2}{1-\tilde{r}^2}
d\Omega_{d-2}^2~,
\end{eqnarray}
which takes a form as in (\ref{warpedform}). With the formula of
Eq.~(\ref{HEE}), the holographic entanglement entropy
functional becomes
\begin{eqnarray}\label{eefunctional}
{\cal S} &=& \frac{L^{2d-3}}{4 G_N^{(d+1)}} \Omega_{d-2} \int_0^{\tilde{R}} \frac{d\tilde{r}}{z^{d-1}}
 Q P^{d-2} \left[1+\frac{2 \lambda z^2}{L^2}\left(P^{-2} +L^2 Q^{-2} \left(\frac{P'}{P}-\frac{z'}{z}\right)^2\right)\right]~.\\
Q &\equiv& \sqrt{\frac{L^2}{(1-\tilde{r}^2)^2}-f(z,v) v'^2 - 2 v'
z'}~, \qquad P \equiv
\frac{\tilde{r}}{\sqrt{1-\tilde{r}^2}}~.\nonumber
\end{eqnarray}
As done in the above two subsections, we need to solve the two
equations of motion to get the extreme value of the holographic
entanglement entropy functional. The two equations of motion are
very complicated and we do not show them explicitly here. We adopt
the same boundary conditions of (\ref{boundarycondition}) in doing
the computation.

 For convenience, we define
the relative holographic entanglement entropy $\delta \bar{{\cal
S}} \equiv \frac{{\cal S}-{\cal S}_{thermal}}{V_b}$, where $V_b =
L^{d-1} \Omega_{d-2} \int_0^{\tilde{R}}
\frac{\tilde{r}^{d-2}}{(1-\tilde{r}^2)^{d/2}} d\tilde{r}= L^{d-1}
\Omega_{d-2} \frac{\tilde{R}^{d-1}}{d-1} ~_2
F_1\left(\frac{d-1}{2},\frac{d}{2},\frac{d+1}{2},\tilde{R}^2\right)$
is the volume of the entangled region $A$ on the boundary. In Fig.
10, we plot the time evolution of $\delta \bar{{\cal S}}$ with
various Gauss-Bonnet coupling constant $\lambda$ for chosen size
$\tilde{R}$ of the entangled sphere. The disclosed saturation time
decreases as $\lambda$ increases, which is consistent with the
results of the previous two observables. Comparing with the
previous two observables, we observe that the delay time in the
onset of thermalization becomes shorter as shown in the
entanglement entropy. This can be understood, since the
entanglement entropy is related more to the degrees of freedom of
the system so that it is more sensitive to the thermalization
process.

Looking at the initial absolute value of $\delta \bar{{\cal S}}$,
we surprisingly find that with the increase of $\lambda$, it
becomes smaller, instead of becoming bigger. This means that the
initial state of the field system due to the higher curvature
correction terms in the gravity is closer to the thermal
equilibrium state, which is completely different from the results
of the previous two observables. The relation of the
thermalization process to the Gauss-Bonnet factor holds as well
when we fix the spacetime dimension to be  $d=5$ as shown in
Fig.~11.

In Fig. 12, we show the spacetime dimensional influence on the
thermalization process reflected by the entanglement entropy. It
is clear that with the  increase of the spacetime dimension, the
initial absolute value of $\delta \bar{{\cal S}}$ increases, while
the saturation time decreases. This is displayed more clearly in
Fig. 13, where the saturation time $\tilde{t}_{sat}$ versus the
size of the entangled sphere is plotted. Comparing with the other
two observables, the entanglement entropy reflects more clearly
the dependence of the saturation time on the spacetime dimensions,
especially for large $\tilde{R}$. This is expected, since the
two-point function and the Wilson loop are only related to the
degrees of freedom on the two points and the loop respectively,
while the entanglement entropy is related to the degrees of
freedom in the volume of the bounded region $A$ which depends
strongly on the spacetime dimensions. This point can also be seen
by comparing definitions of these three observables in Eqs.~(\ref{lengthfunctional}),(\ref{areafunctional}) and
(\ref{eefunctional}). From Fig. 13, it is clearly shown that the
saturation time decreases as $d$ increases. Physically, we can
understand this phenomenon, since quanta in higher dimensions have
more degrees of freedom to collide and interact with neighboring
quanta, and thus thermalize faster.

\begin{figure}[!htbp]
\centering
\includegraphics[width=0.3\textwidth]{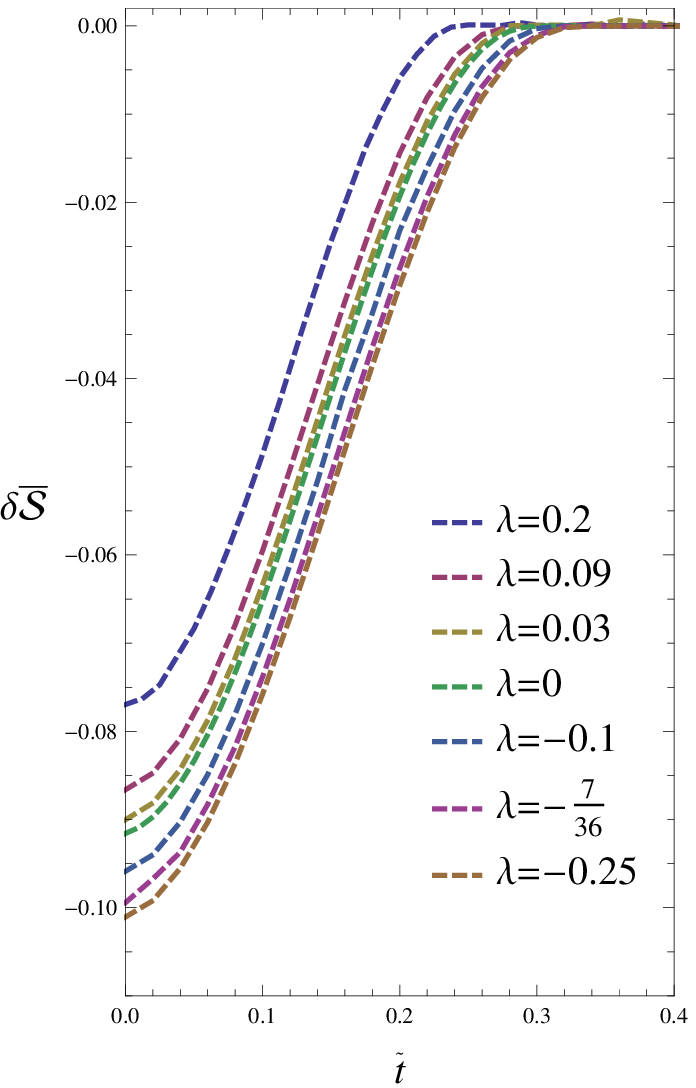}\qquad\qquad
\includegraphics[width=0.29\textwidth]{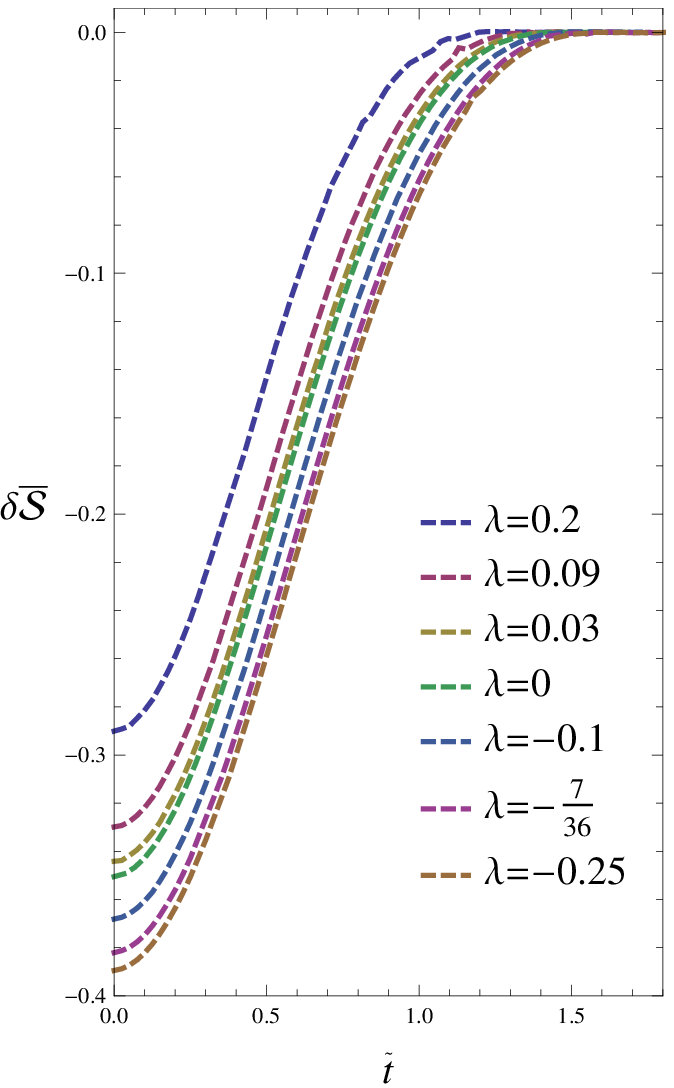}
\caption{Time evolution of $\delta \bar{{\cal S}}$ in $d=4$ case.
From left to right, the rescaled radius of the entangled sphere
are $\tilde{R}=0.3, 0.9$, respectively. In each figure, the
Gauss-Bonnet coupling constant $\lambda$, from top to bottom, are
taken to be $0.2, 0.09, 0.03, 0, -0.1, -\frac{7}{36}, -0.25$.}
\end{figure}

\begin{figure}[!htbp]
\centering
\includegraphics[width=0.3\textwidth]{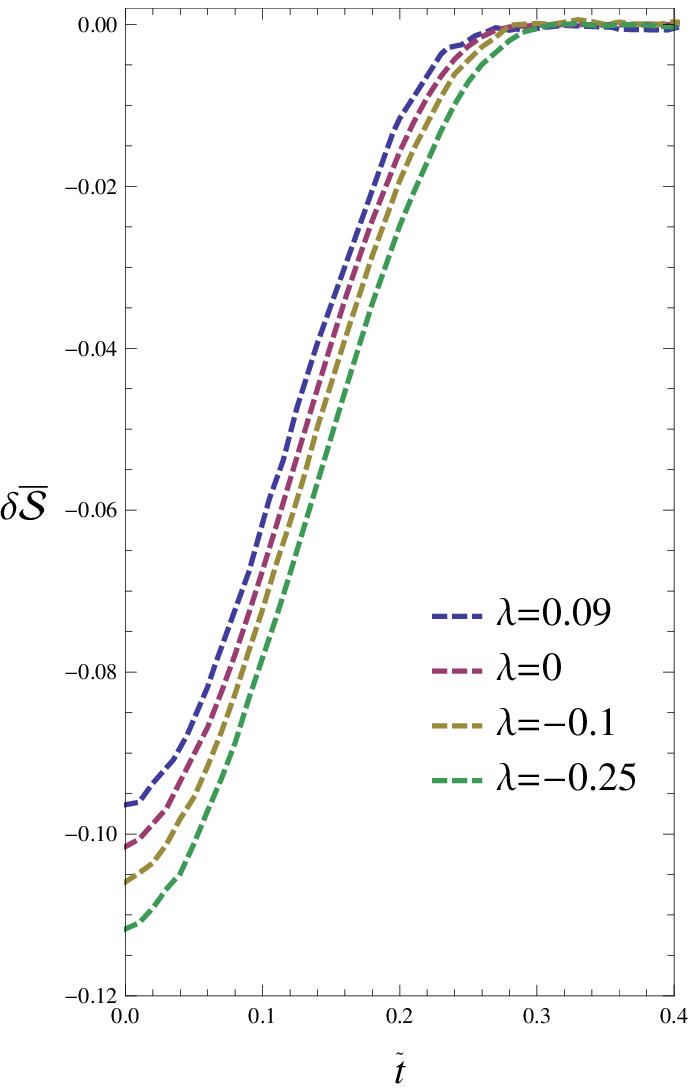}\qquad\qquad
\includegraphics[width=0.29\textwidth]{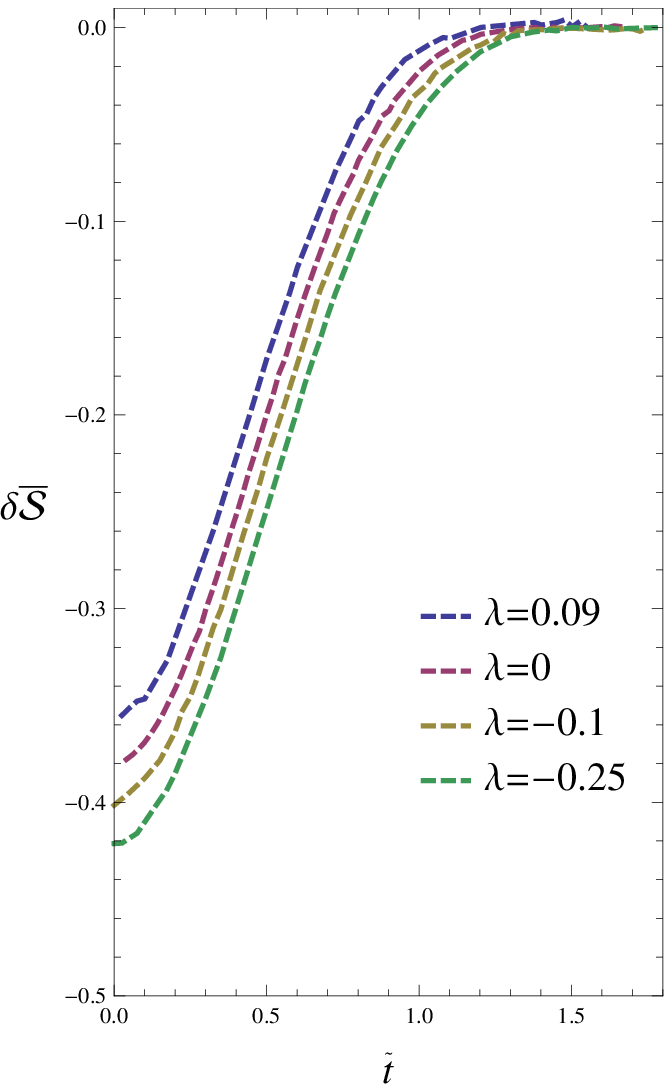}
\caption{Time evolution of $\delta \bar{{\cal S}}$ in $d=5$ case,
the rescaled radius of the entangled sphere $\tilde{R}=0.3, 0.9$,
respectively. In each figure, the Gauss-Bonnet coupling constant
$\lambda$, from top to bottom, are taken to be $0.09, 0, -0.1,
-0.25$.}
\end{figure}

\begin{figure}[!htbp]
\centering
\includegraphics[width=0.3\textwidth]{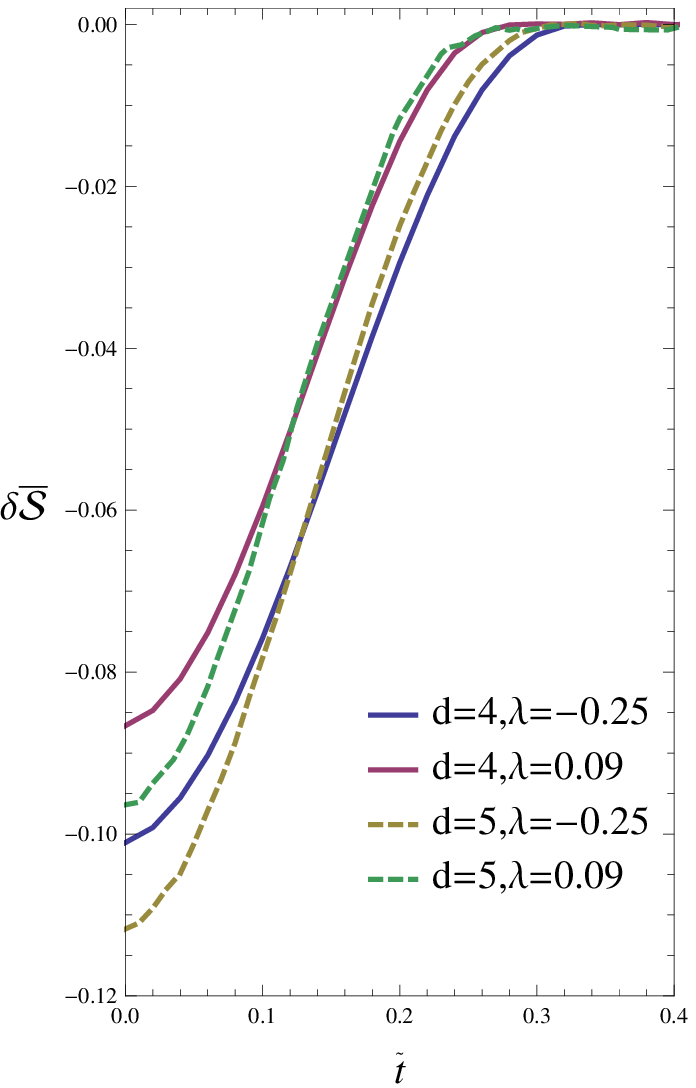}\qquad\qquad
\includegraphics[width=0.288\textwidth]{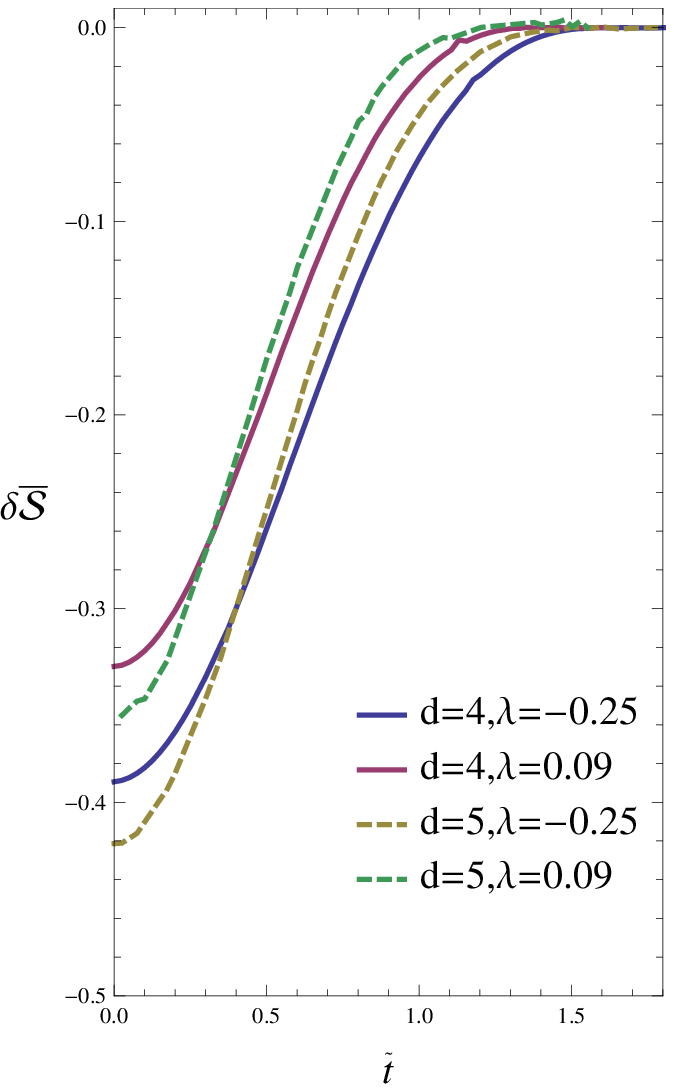}
\caption{Time evolution of $\delta \bar{{\cal S}}$ in $d=4$ and
$d=5$ cases. From left to right, The rescaled radius of the
entangled sphere are $\tilde{R}=0.3, 0.9$ respectively.}
\end{figure}

\begin{figure}[!htbp]
\centering
\includegraphics[width=0.49\textwidth]{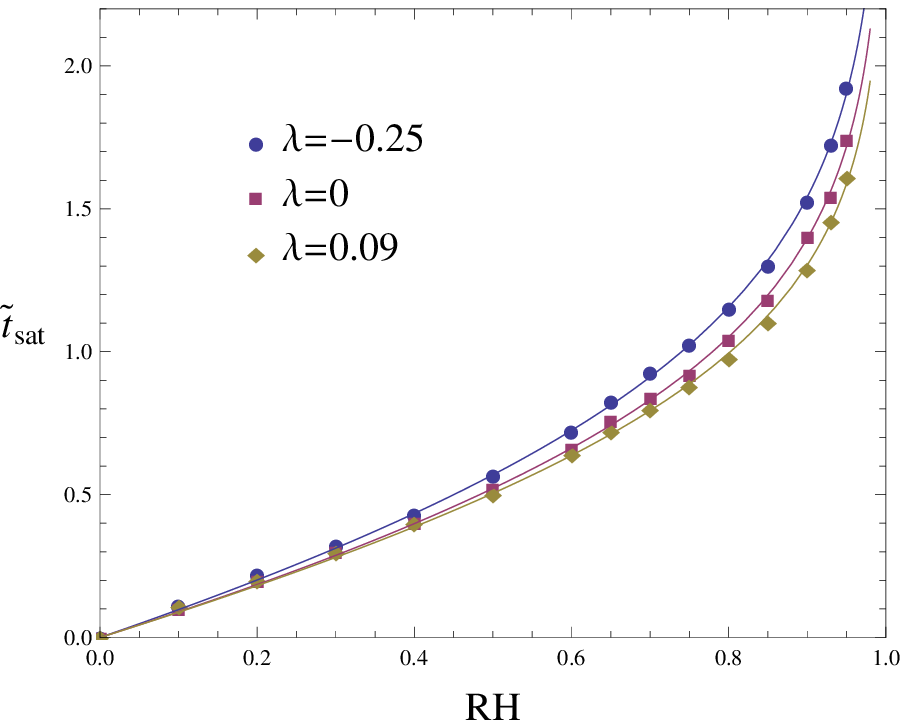}
\includegraphics[width=0.49\textwidth]{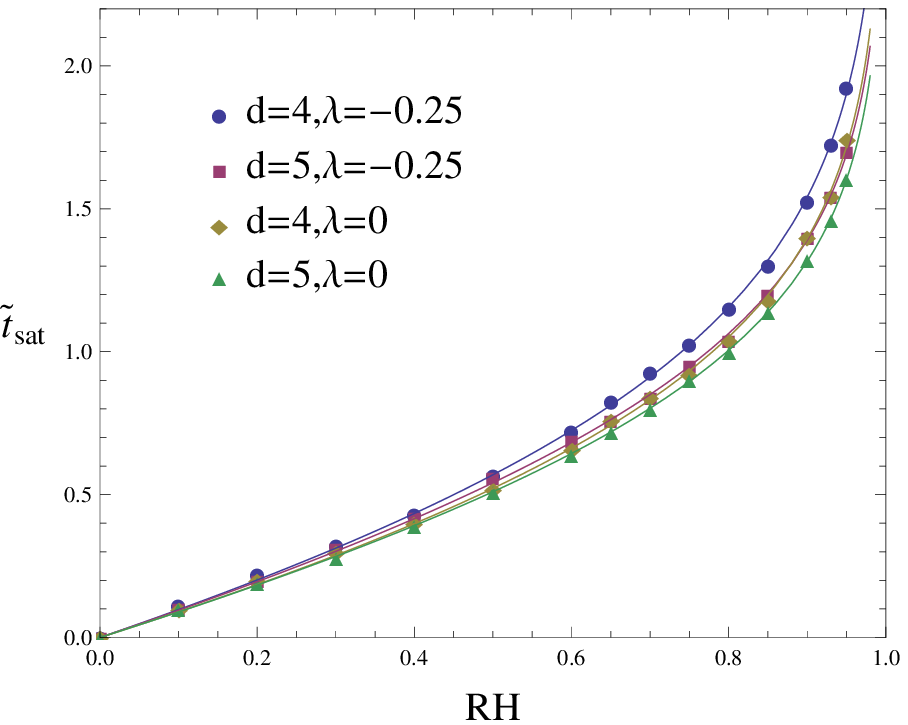}
\caption{Saturation time $\tilde{t}_{sat}$ as a
function of the boundary scale $\tilde{R}$. The
left panel is for $d=4$. The right panel is the
comparison of the dimensional influence by fixing
the Gauss-Bonnet factor. Solid curves in both panels are produced by fitting the numerical points with function $a \tilde{R} + b \ln (1-\tilde{R})$. There is a large overlap between the curve with $(d=4, \lambda=0)$ and the one with $(d=5, \lambda=-0.25)$.}
\end{figure}

\section{Conclusions and discussion}

In this work we have discussed the holographic thermalization
process by relating gravity theory with higher curvature
corrections to the dual  strongly coupled quantum field theory
living on the dS boundary. We found that the whole thermalization
process follows the general pattern observed in
\cite{Fischler:2013fba}. At very early time the evolution of the
thermalization encounters a delay, then it enters the
pre-local-equilibration stage during which the growth is quadratic
in time. Afterwards we have the post-local-equilibration stage of
linear growth, and finally there emerges a period of memory loss,
after which the curves flatten out and the observables reach their
thermal values.

Our findings follow some universal properties, as observed in the
Einstein case with flat
\cite{Liu:2013iza,Liu:2013qca,Balasubramanian:2010ce,Balasubramanian:2011ur}
or with curved \cite{Fischler:2013fba} boundary, but also they
show some district features due to the presence of higher
curvature correction term in the bulk. As we discussed in the
Introduction, a  delay in the onset of thermalization was found.
In our case for large boundary scales $\tilde{R}$ we also observed
a delay on the onset of thermalization for the two probes, the
two-point function and the Wilson loop, while the onset of
thermalization becomes much shorter for the third probe, the
entanglement entropy.

In Einstein gravity the saturation time, i.e the time needed the
system to reach thermal equilibrium is very fast. In our study for
larger boundary scales $\tilde{R}$ it takes more time for the
system to reach thermal equilibrium and this result holds for all
the three probes. However, as the strength of the Gauss-Bonnet
coupling $\lambda$ is increased, the saturation time of the
thermalization process  becomes shorter as the entanglement
entropy shows. This is interesting because it relates directly the
straight of gravity effects to the time that the system on the
boundary needs to reach thermal equilibrium.

In Einstein gravity the thermalization process seems to be
independent of dimensionality. In the presence of the higher
curvature correction term in the bulk the two probes the two-point
function and the Wilson loop show the same behavior. However, the
entanglement entropy shows clearly a spacial dependence of the
thermalization process. This can be attributed to the fact that
the holographic entanglement entropy contains more bulk
information and is more sensitive to the spacetime dimensions. We
have  found that the thermalization process as probed by the
entanglement entropy is faster in the spacetime with higher
dimensions. This can be understood since quanta in higher
dimensions have more degrees of freedom to collide and interact
with neighboring quanta, which makes the thermalization quicker.

It is interesting to investigate further the effect of a
nontrivial gravity bulk on the thermalization process. A way to
study this effect is to consider a scalar field collapsing in an
AdS space. Following this approach it was found
\cite{danielsson,janik1,chesler,garfinkle,Garfinkle:2011tc} that
for  a collapsing scalar field minimally coupled to gravity the
thermalization process proceeds very rapidly. One can introduce a
nonminimal coupling like a coupling of a scalar field to Einstein
tensor and investigate the thermalization process on the boundary.
Work in this direction is in progress.

\section*{Acknowledgments}

We thank Shao-Feng Wu and Yong-Zhuang Li for
their kind help on the numerical calculations.
This work is supported in part by the National
Natural Science Foundation of China. E. A. thanks the support of FAPESP and CNPq (Brasil). E.P. is
partially supported by the ARISTEIA II Action of the Operational
Programme "Education and Lifelong Learning" which is co-funded by
the European Union (European Social Fund) and National
Resources.

\end{document}